\documentclass[12pt]{iopart}

\usepackage{iopams}
   \usepackage[dvips]{graphicx}

\newcommand{\apjl}{\emph{ApJL}}
\newcommand{\apj}{\emph{ApJ}}
\newcommand{\aap}{\emph{A\&A}}
\newcommand{\aj}{\emph{AJ}}
\newcommand{\nat}{\emph{Nature}}
\newcommand{\mnras}{\emph{MNRAS}}

\begin{document}


\title{Methane in the atmosphere of the transiting hot Neptune GJ436b?}


\author{J.-P. Beaulieu$^{1,2}$, G. Tinetti$^{2}$, D. M. Kipping$^{2,3}$,
I. Ribas$^{4}$, R. J. Barber$^{2}$, J. Y-K. Cho$^{5}$, I. Polichtchouk$^{5}$,
J. Tennyson$^{2}$, S. N. Yurchenko$^6$, C. A. Griffith$^7$, V. Batista$^{1}$, I. Waldmann$^{2}$, S. Miller$^{2}$, 
S. Carey$^8$, O. Mousis$^9$, S. J. Fossey$^{2}$, A. Aylward$^{2}$}

\address{ $1$ Institut d'Astrophysique de Paris, UMR7095, CNRS, Universit\'e Paris VI, 98bis Boulevard Arago, 75014 Paris, France \newline
$^2$ Department of Physics and Astronomy, University College London, Gower street, London WC1E 6BT, UK  \newline
$^3$ Harvard Center for Astrophysics, 60 Garden Street, Cambridge, USA \newline
$^4$ Institut de Ciencies de l'Espai (CSIC-IEEC), Campus UAB, 08193 Bellaterra, Spain \newline
$^5$ Astronomy Unit,  Queen Mary University of London, Mile End Road, London E1 4NS, UK \newline
$^6$ Institut fur Physikalische Chemie und Elektrochemie, Technische Universitat Dresden, D-01062 Dresden, Germany \newline
$^7$ Lunar and Planetary Laboratory, University of Arizona, Tucson, AZ, US \newline
$^8$ IPAC-Spitzer Science Center, California Institute of Technology, Pasadena,  91125, CA, USA \newline
$^{9} $ Universit{\'e} de Franche-Comt{\'e}, Institut UTINAM, CNRS/INSU, UMR 6213, 25030 Besan\c{c}on Cedex, France}
\ead{beaulieu@iap.fr}


\begin{abstract}

We present an analysis of seven primary transit observations of the hot Neptune GJ436b at 3.6, 4.5 and $8~\mu$m obtained with the Infrared Array Camera (IRAC) on the Spitzer Space Telescope.  After correcting for systematic effects, we fitted the light curves  using the Markov Chain Monte Carlo technique. Combining these new data with the EPOXI, HST and ground-based $V, I, H$ and $K_s$  published observations, the range  $0.5-10~\mu$m  can be covered. Due to the low level of activity of GJ436, the effect of starspots on the combination of transits at different epochs is negligible at the accuracy of the dataset.
Representative climate models were calculated by using a three-dimensional, pseudo-spectral general circulation model with idealised thermal forcing. 
 Simulated transit spectra of GJ436b were generated using line-by-line radiative transfer models including the opacities of the molecular species expected to be present in such a planetary atmosphere. A new, ab-initio calculated, linelist for hot ammonia has been used for the first time. 
The  photometric data observed at multiple wavelengths can be interpreted with methane being the dominant  absorption after molecular hydrogen, possibly with minor contributions from ammonia, water and other molecules. No clear evidence of carbon monoxide and dioxide is found from transit photometry. We discuss this result in the light of a recent paper where photochemical disequilibrium is hypothesised to interpret secondary transit photometric data. We show that the emission photometric data are not incompatible with the presence of abundant methane, but further spectroscopic data are desirable to confirm this scenario.
\end{abstract}





\section{Introduction}
{ As the closest transiting planet to date, GJ436b is the smallest and coolest ($\sim 700$K) exoplanet for which both transmission and emission flux can be measured at optical to IR wavelengths. This hot
Neptune transits a nearby bright ($K_s=6.07$) M2.5V dwarf star at 0.029 AU with a period $\sim$2.6438986 days (Butler et al. 2004; Gillon et al. 2007,  Demory et al., 2007).  Although the planet is small with a transit depth of $\sim 0.7 \%$, its atmospheric temperature and planetary parameters (Deming et al. 2007) indicate an extended atmosphere, making it an excellent candidate for the detection of atmospheric {constituents} using the primary transit technique. Thus GJ436b 
provides a unique opportunity to extend investigations of exoplanetary atmospheres to smaller Neptune-mass planets. 
Recent studies indicate the distinct nature of GJ436b's orbit and composition.  The exoplanet's mass and size reveal 
a body that is denser and thus of different internal structure than the Jovian-sized planets. The orbital parameters indicate an 
somewhat eccentric orbit; thus the planet is probably not tidally locked, like most hot-Jupiters studied so far (Nettlemann et al. 2010). 
The composition of GJ436b, as investigated with secondary eclipse observations of GJ436 at 3.6, 4.5, 5.8, 8, 16 and 24 $\mu$m
(Stevenson et al. 2010), indicates an atmosphere containing a high abundance of CO, some  water and a low abundance of methane.
This composition is far out of equilibrium chemistry, for which models at the relevant temperatures suggest that both water and methane should be present and relatively abundant in the atmosphere of this hot-Neptune (Sharp and Burrows, 2007; Lodders et al. 2002), possibly together with NH$_3$,  hydrocarbons and H$_{2}$S.  Thus Stevenson et al. (2010) postulate the presence that two disequilibrium 
mechanisms: vertical mixing to bring CO from deeper and warmer levels where it is abundant in equilibrium, and photochemistry to destroy methane and thus explain its low abundance.   
Here we present an analysis of primary transit photometry at 3.6, 4.5 and $8~\mu$m, which suggests, in contrast with Stevenson et al.'s (2010) study of the secondary transit photometry data, that CH$_4$ is the most abundant 
carbon molecule.  In order to more fully examine the composition of GJ436b's atmosphere, we also reanalyzed the secondary transit 
data.  We derived fluxes that confirm those obtained by Stevenson et al. (2010). However we obtained larger errors 
for the measured fluxes at 3.6 and 4.5 $\mu$m, which allow for a composition that is also CH$_4$-rich, consistent with our analysis 
of the primary transit data.   In the following sections, we discuss an analysis of 
new primary transit data, the re-analysis of previously published secondary transit data, and examples of radiative transfer interpretations 
that fit both transit data with CH$_4$ as the dominant carbon molecule, consistent with thermochemical equilibrium solutions.  }

\section{The photometric light curves}

Seven primary transits of GJ436b have been observed by Spitzer IRAC as part of two Spitzer programs (40685 and 50051).  Two epochs have been obtained  at each of 3.6~$\mu$m and 4.5$~\mu$m, and three epochs at 8~$\mu$m.  Unfortunately, no data were obtained at 5.8 $\mu$m before the end of the cryogenic life of Spitzer. Observations have been carried out in sub-array mode, and images are delivered in cubes of 64 slices of 32x32 pixels. We median-stack all the slices within a cube, and then process the data using the method described { in
the section 2.2 of }  Beaulieu et al. (2010) for HD 209458b. Note that the data set is very similar, the target being of similar brightness in the infrared.

\begin{figure*}
\begin{center}
\includegraphics[angle=0,width=8 cm]{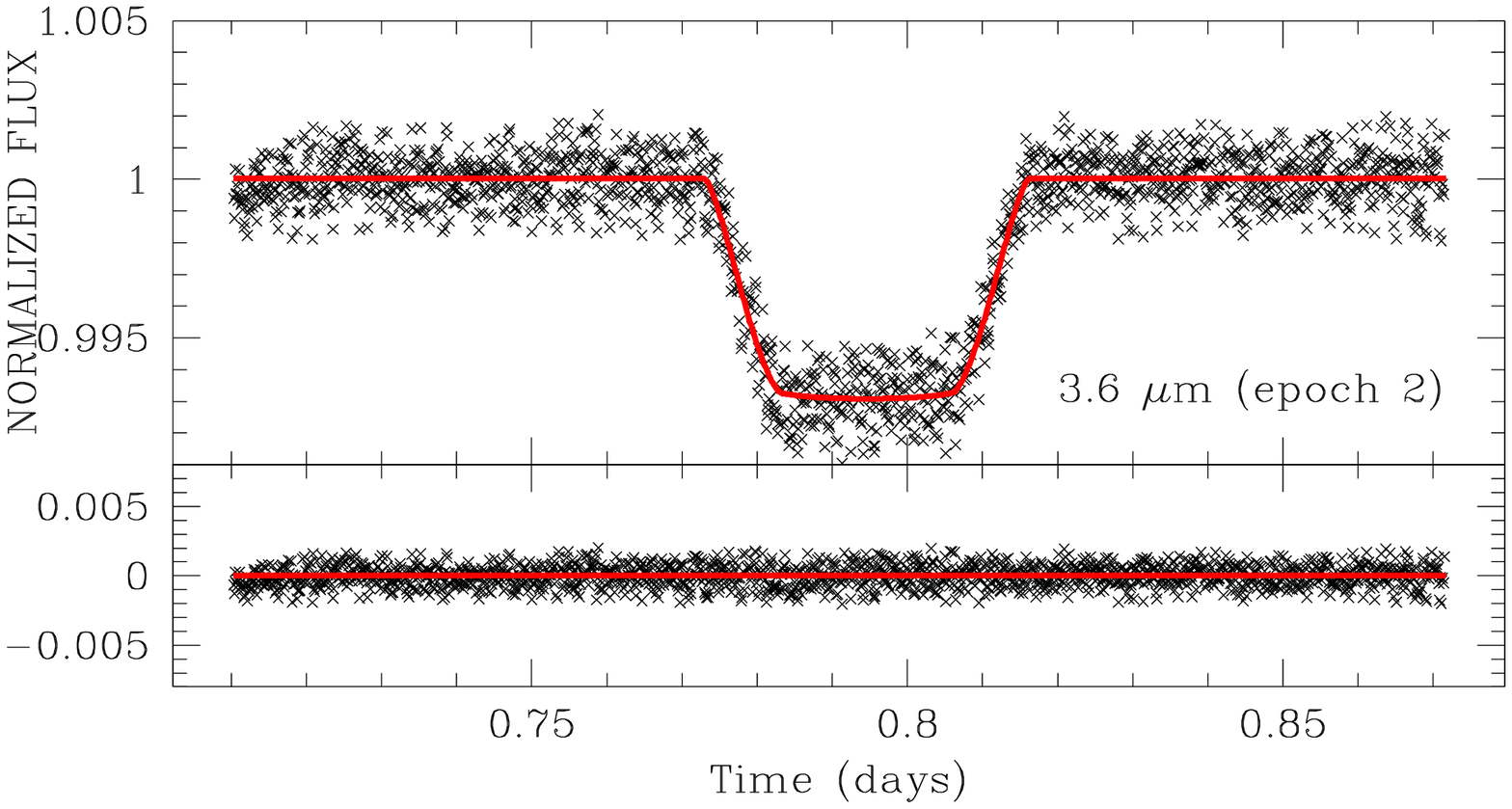}\includegraphics[angle=0,width=8. cm]{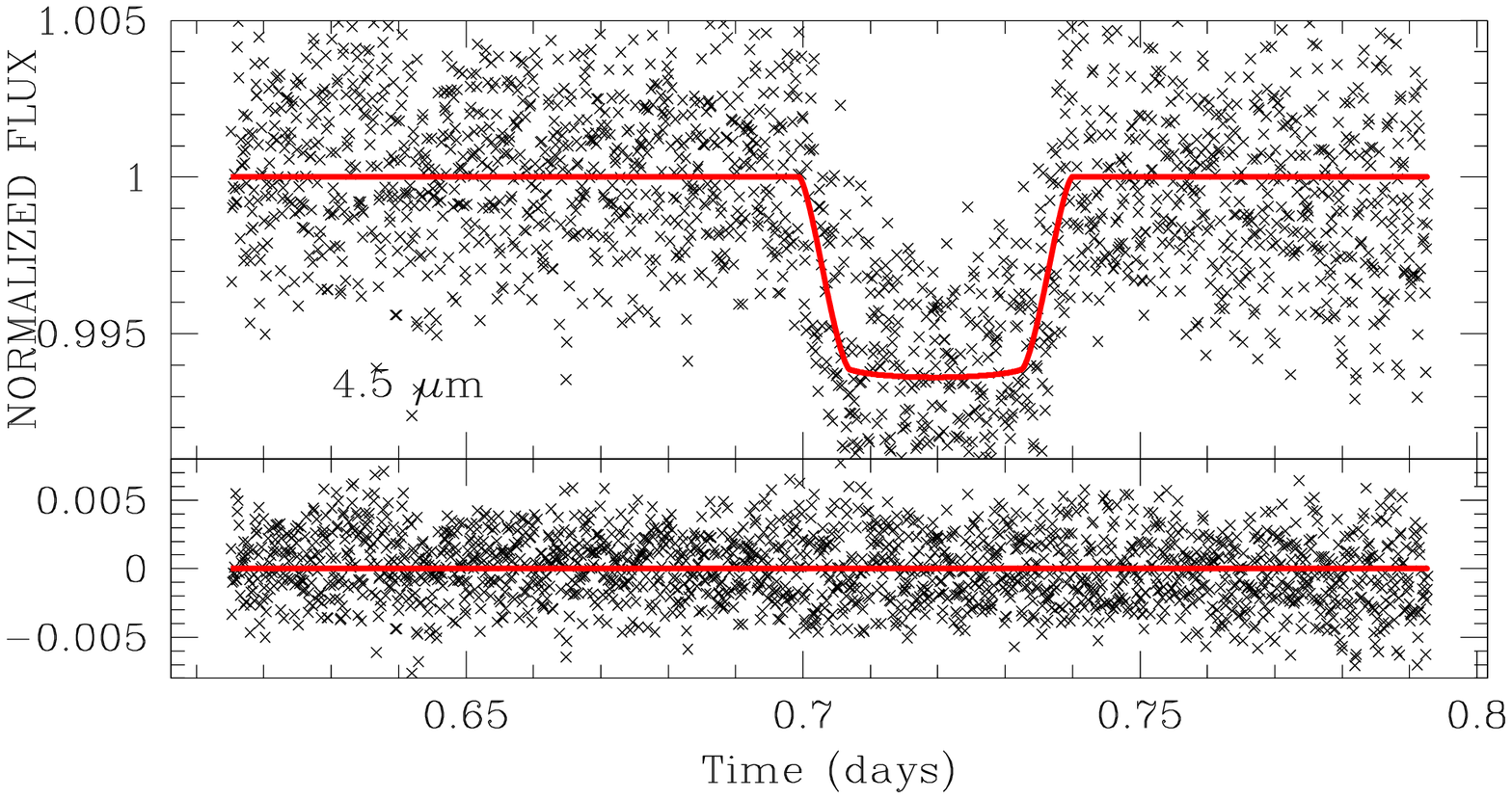}
\includegraphics[angle=0,width=8. cm]{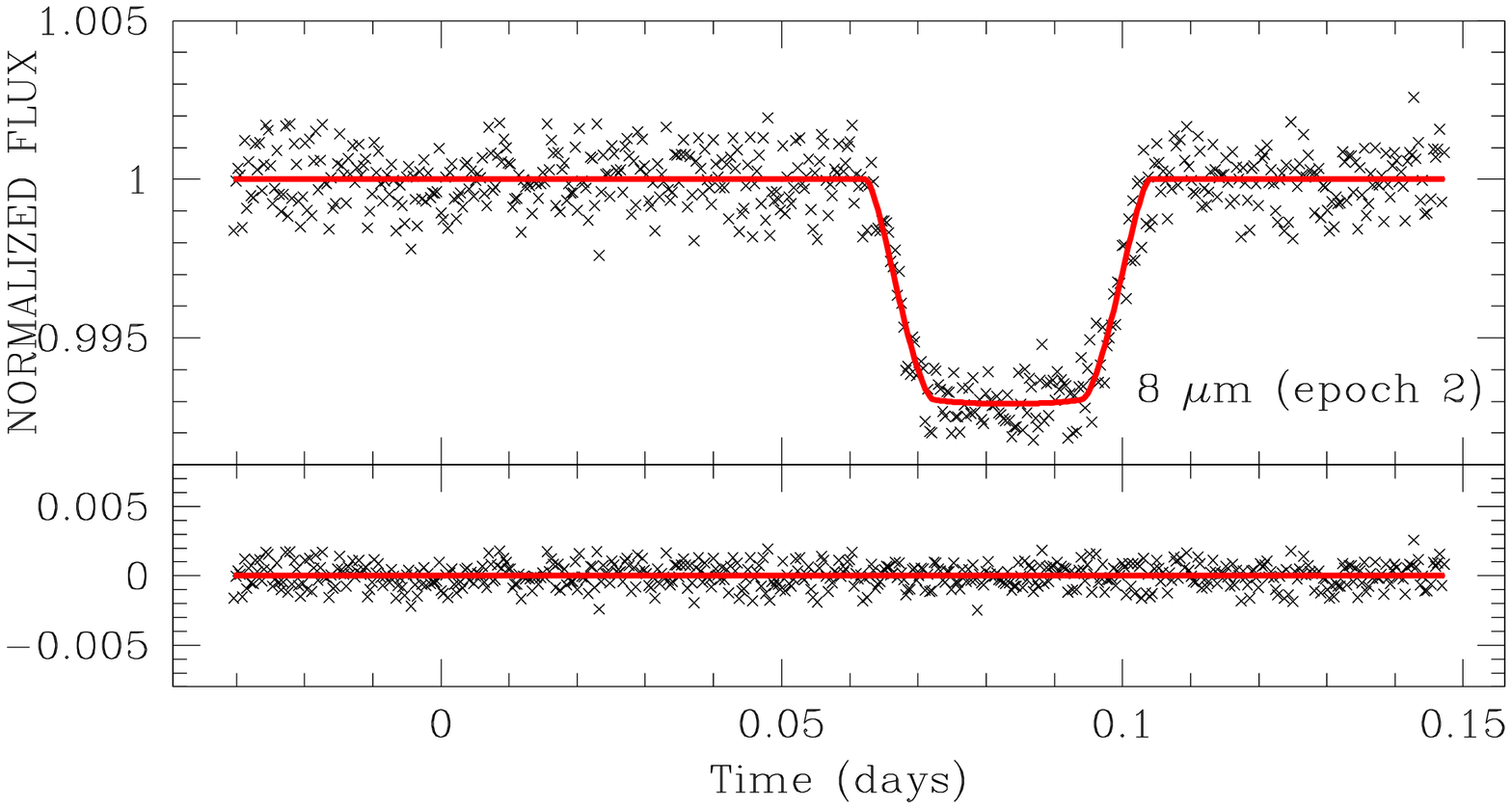}\includegraphics[angle=0,width=8. cm]{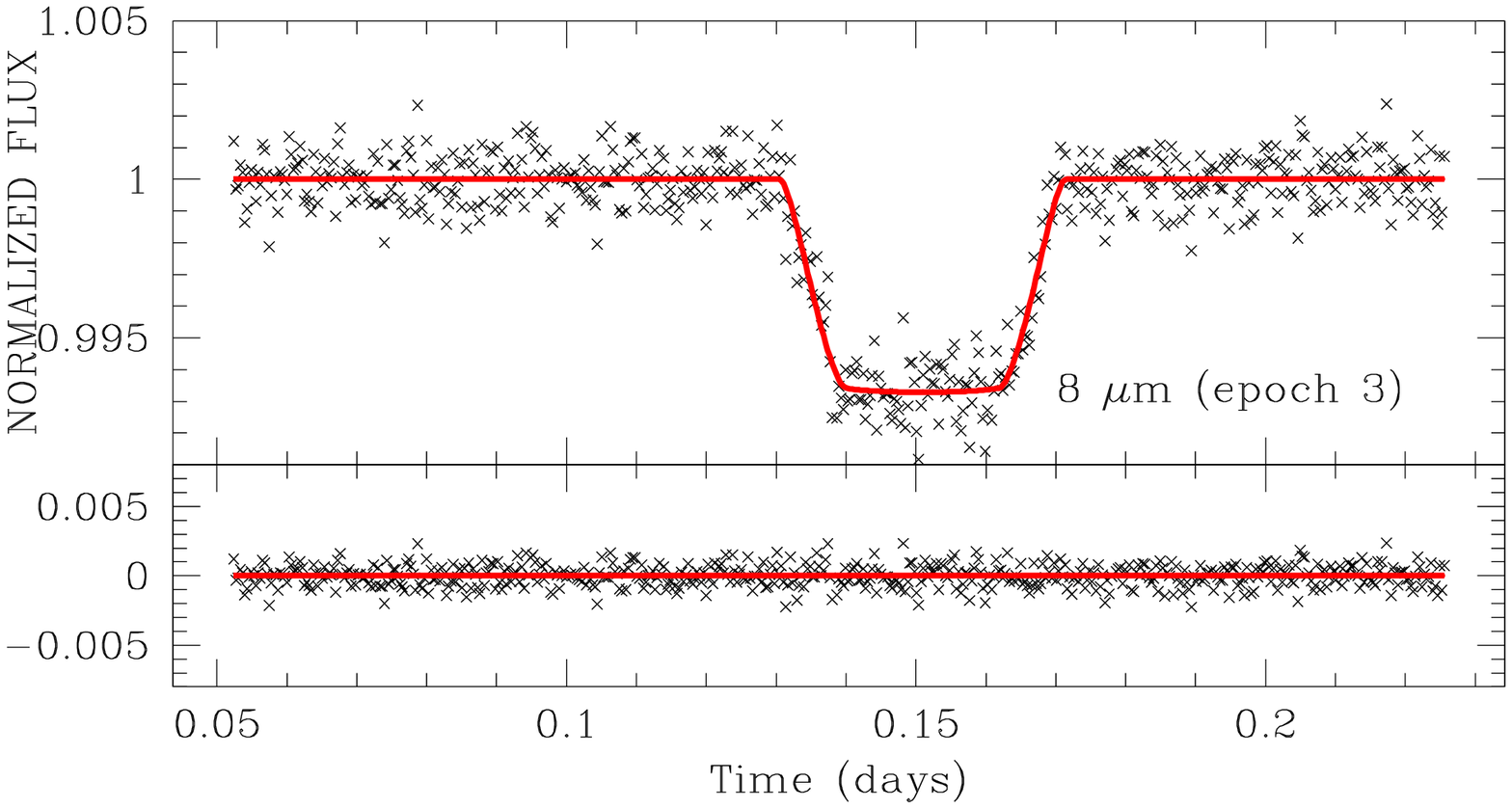}
\caption{Final light curves, best fit model and residuals at 3.6, 4.5 and the two most recent epochs of 8 $\mu$m. The first 
epoch at 8 $\mu$m has been published by Deming et al. (2007) so we do not show here our re-analysis. Raw data are shown in the
appendix in Figure~\ref{fig:lc3.6-4.5}. } \label{fig:lcfinal}
\end{center}
\end{figure*}

Spitzer's IRAC instrument has four bandpasses split into two types of detector, which display very different instrumental systematics. The 3.6~$\mu$m and 4.5~$\mu$m channels are indium/antimony detectors, which are known to exhibit ostensibly periodic flux variations. These variations are a result of a non-uniform response function within each pixel, combined with small pointing variations of the Spitzer spacecraft (see Morales-Calderon et al. 2006; Beaulieu et al. 2008). Recently, it has been proposed that these bandpasses also exhibit a time dependence, as well as pixel phase variations (S2010).

The 5.8~$\mu$m bandpass (for which no observations of primary transit of GJ436b exist) and 8.0~$\mu$m bandpasses are arsenic-doped silicon chips exhibiting strong temporal-ramp effects, the source of which is believed to be charge trapping (IRAC data handbook 2006). The 8.0~$\mu$m channel is arguably the simplest to   correct of all four bandpasses. 

Due to the strong dependence of the transit depth on an accurate correction for systematic effects, the full details of our corrective procedure will be described and discussed in the appendix.  We summarise 
a few points here: 

 The time scales and amplitudes of the primary transit of GJ436b and of the systematics due to pixel phase at 3.6~$\mu$m and 4.5~$\mu$m are very similar.  As a consequence, contrary to other observations where several cycles of pixel phase are present in and out of transit, the mutual phasing of the astrophysical signal and the systematics becomes a critical issue. We carefully corrected for systematics and checked for the stability of the corrections  applied to the data; we derived the final values at 3.6, 4.5 and 8 ~$\mu$m to be used in the analysis (Table 1). 

Concerning the S2010 reduction of secondary transit observations, we obtained similar results at 5.8~$\mu$m and 8~$\mu$m while   we  find greater discrepancy with the measurements at 3.6 and 4.5 ~$\mu$m. The nature of the systematics at 3.6 and 4.5  ~$\mu$m are similar to our primary transit observations. The 3.6  ~$\mu$m secondary transit has a mutual phasing of the astrophysical signal and the systematics close to 0, which is the worst case scenario.
It is a similar situation to the first epoch of our primary transit at 3.6 ~$\mu$m: in that case we had two epochs, though, so we could
discard the one with such mutual phasing as unreliable. Here there is no such possibility, so we propose  adopting a larger errorbar at 3.6 ~$\mu$m: $~ 0.03 \pm 0.02\%$. At 4.5 ~$\mu$m we measure a transit depth of $0.01\pm 0.01$ \% in contrast with the $3\sigma$ upper limit of $0.01$ \% proposed by S2010.
As the interpretation of the secondary transit photometric data strongly relies on the 3.6 and 4.5 ~$\mu$m
-in particular the hypothesis of photochemical disequilibrium and paucity of methane-
 additional measurements of secondary transits at 3.6 and 4.5 ~$\mu$m bandpasses are critically needed.

\subsection{Stellar activity of GJ436}
{ A possible source of systematic errors when combining data taken over different epochs and in different bands is the influence of magnetic activity, i.e., starspots. This was discussed in detail by Beaulieu et al. (2008) for HD189733. The magnetic activity level of GJ436 can be assessed using a number of indicators. Its X-ray luminosity has been measured to be $L_{\rm x} = 9 \cdot 10^{25}$ erg~s$^{-1}$ (Sanz-Forcada et al. 2010), which is much lower than the value of the Sun of $L_{\rm x} = 2 \cdot 10^{27}$ erg~s$^{-1}$. Even accounting for the $\sim$8 times larger emitting solar surface, the flux density from the corona of GJ436 is still $\sim$3 times smaller than the solar value. The chromospheric Ca H \& K indicator also suggests a very low activity object, in agreement with its low rotation velocity (Jenkins et al. 2009) and with its kinematic properties typical of an old disk star (Browning et al. 2010). Time series data from the ground (Butler et al. 2004) and space (Ballard et al. 2010) do not find evidence for photometric variations with an amplitude above $\sim$0.5 mmag in the visible bands, which is in good overall agreement with the rest of the activity indicators. The only hint of (modest) stellar activity  comes from ground-based time series photometry presented by Demory et al. (2007), which shows scatter with a peak-to-peak amplitude of $\sim$10 mmag possibly caused by rotational modulation. This may indicate the existence of time-dependent photometric variations but the small number of measurements deviating from the mean in the Demory et al. study prevents a more quantitative analysis.

The effect of spot activity can be further studied by measuring variations in the transit depth at different epochs, arising from changes in the spot coverage of the strip occulted by the planet. If such an effect was important, the depth of the transit would vary over time and this could be observed as an additional scatter in a collection of depth measurements. We used the homogeneous list of transit parameters in Coughlin et al. (2008) and calculated the standard deviation of the depth measurements. If we consider the 6 professional measurements in the $V$ band (combining all visible measurements is not adequate given the effect of limb darkening), which cover a timespan of $\sim$400 days, the weighted depth average is $6.73\pm0.20$ mmag and $\chi^2=0.51$. Further, the $I$-band measurements of Ribas et al. (2009) plus several (unpublished) additional ones totaling 8 data points over a 750-day timespan (3 seasons) yield a depth value of $7.02\pm0.17$ mmag, and $\chi^2=1.56$.

The resulting $\chi^2$ values, close to unity, indicate that random noise is the dominant factor and activity noise is absent or at least probably below $\sim$0.02\% in the $V$ and $I$ bands. This can be translated into a { conservative} upper limit of $\sim$0.01\% at 3.6 $\mu$m. Given the overall low activity level of GJ436, the effect of starspots on the combination of transits at different epochs is negligible at the quality of our dataset, well below our typical transit depth measurement error.
}
\section{Results}

\subsection{Analysis of Spitzer primary transits at 3.6, 4.5 and 8 $\mu$m}

We have five high quality primary transit light curves with well understood and corrected systematic effects. Note that the 2007 data at 8 $\mu$m have been published already (Gillon et al. 2007; Deming et al. 2007; Southworth 2008).
In this work we processed all the data in a uniform way and present them together.
First, we calculate accurate limb darkening coefficients for each of the IRAC band following the procedure described by Beaulieu et al. (2010). With the four coefficients  { Claret (2000) parameterization for limb darkening },  we obtained (0.7822,  -0.8644, 0.5827, -0.1557) for 3.6 $\mu$m,  (0.6087, -0.5608, 0.3510, -0.0918)  for 4.5 $\mu$m and  (0.5727, -0.6246, 0.4055, -0.1026) for 8 $\mu$m.

We adopted the physical model of a transit light curve following the expression of Mandel \& Agol (2002) and orbital eccentricity following the equations of Kipping (2008). We sampled the parameter space with a Markov Chain Monte Carlo code. We adopted a fixed value of period $P=2.6438986$ days (Ballard et al. 2010). For each light curve, 5 parameters were fitted, namely the out-of-transit baseline, the orbital inclination $i$, the ratio between the orbital semi-major axis and the stellar radius $a/R_*$, the ratio of radii $k$ and the mid transit time $t_c$. The results are shown in Table 1. The best fit model and residuals for each channel are plotted in Figure \ref{fig:lcfinal}. Note that we obtained a slightly shallower eclipse depth at 8~$\mu$m than Deming et al. (2007), because of the different treatment of the ramp correction, i.e. a polynomial fit versus the Agol et al. (2009) approach, but it is compatible within the error bar. {  After submission of the present article, Agol et al. (2010) proposed a new approach with a double exponential function is physically motivated with a proper asymptotic behavior perfectly adapted also for long time time series obtained for phase curves. The results of the different approachs are compatible within their error bars.}

\subsection{Final values for primary transits to be used for the analysis}

We considered additional high accuracy ground based and space based measurements.
In particular C\`aceres et al. (2009) measured the transit depth of GJ436b in $K_s$ band in May 2007 to be $0.64 \pm 0.03 \%$, while Alonso et al. (2008) reported  $0.697 \pm 0.023 \%$  in $H$ band. From HST observations  in November-December 2007, the mean transit depth between $1.35-1.85~\mu$m was estimated to be  $0.691 \pm 0.012 \%$ (Pont et al. 2008). Ballard et al. (2009) combining all the data collected by the EPOXI mission obtained a  transit depth of $0.663 \pm 0.014 \%$ over a broad band of $0.35-1.0~\mu$m; this data set is not public yet. We note though that Ballard et al. (2009) performed a spline fit to the out-of-transit data before measuring the transit depth: as the data are affected by systematics with a similar time scale to the transit, 
this procedure may affect the derived results.
 We considered also the combined ground based $V$ and $I$ band measurements mentioned in the previous section:   
in $I$ the transit depth obtained was $0.702\pm0.017 \%$ and in $V$, $0.673\pm0.020 \%$ (Coughlin et al., 2008; Ribas et al., 2009).
\begin{table*}
{\tiny
\begin{tabular}{l l l l l  l l l } 
\hline 
\bf Band & \boldmath$p^2=(R_p/R_*)^2$ & \boldmath$p=R_p/R_*$ & \boldmath$T_c$\bf (day) &  \boldmath $T$\bf (s) & \boldmath$a/R_*$ & \boldmath $b$ & \boldmath $\Upsilon/R_*$\bf (days$^{-1}$)  \\
\hline 
3.6 $\mu$m & $0.712\pm0.006 \% $ & $0.08439\pm0.00035$& $2454859.79494\pm0.00008$ &  $2937 \pm 15$ & $14.21_{-0.30}^{+0.30}$ &  $0.8471_{-0.073}^{+0.071}$ & $58.83 \pm  0.30$  \\
\hline
4.5 $\mu$m & $0.638 \pm 0.018 \% $    & $0.07988 \pm 0.0012 $   &  $2454850.54169 \pm 0.00023$    &  
$2907 \pm 45$ & $15.44_{-1.10}^{+1.23}$ &  $0.794_{-0.034}^{+0.043}$   & $59.44_{-0.91}^{+0.94}$ \\
\hline 
8.0 $\mu$m & $0.6847 \pm  0.012 \%$ & $0.08275_{-0.00074}^{+0.00075}$  &$2454850.54169 \pm 0.00021$      & $2858  \pm 31$  & $13.34_{-0.45}^{+0.48}$ & $0.856_{-0.012}^{+0.01}$   & $60.46_{-0.67}^{+0.68}$ \\ 
8.0 $\mu$m & $0.675 \pm 0.012  \%$ & $0.08219_{-0.00071}^{+0.00071}$  &  $2454856.65119 \pm 0.00015$      & $2793  \pm 29$  & $14.38_{-0.54}^{+0.59}$ & $0.839_{-0.015}^{+0.013}$   & $61.88_{-0.64}^{+0.65}$ \\
8.0 $\mu$m & $0.715 \pm 0.013  \%$ & $0.08455_{-0.00075}^{+0.00075}$  &  $2454864.58340 \pm 0.00016$      & $2835  \pm 31$  & $14.07_{-0.64}^{+0.64}$ & $0.841_{-0.015}^{+0.016}$   & $60.94_{-0.66}^{+0.67}$ \\
\hline  
 \multicolumn{7}{c}{ {\it {\bf 8 $\mu$m  combined}} }\\ 
\hline  
8.0 $\mu$m & $0.6921 \pm 0.0072  \%$ & $0.08319_{-0.00043}^{+0.00043}$  &  
      & $2827  \pm 18$  & $13.84_{-0.32}^{+0.33}$ & $0.8475_{-0.008}^{+0.0074}$   & $61.11_{-0.39}^{+0.39}$ \\
\hline 
\end{tabular} }
\caption{\emph{Best-fit transit depths, ratio of radii $p$, mid transit time $T_c$ in BJD (UTC), duration $T$, orbital semi-major axis divided by 
the stellar radius $a/R_*$, inclination $i$ and  $\Upsilon/R_*$  found using the Markov Chain Monte Carlo 
fit method described in \S4.1}} 
\label{table:david} 
\end{table*}

\subsection{Temperature distribution for GJ436b}
{  The effective temperature of GJ436b is $\sim$649 K, assuming that the insolation uniformly
heats the entire planet (Lewis et al. 2010).  However, 
the vertical temperature profile, as well as its variation across GJ436b's disk, 
is still highly unconstrained, due to that paucity of spectroscopic data.  
Lewis et al. (2010) explored the effects of the opacity and eccentricity on the general circulation and heat distribution
with a 3D general circulation model (GCM) that includes the radiative transfer effects due to gaseous absorption. They find that 
the departure of GJ436b's eccentricity from a circular orbit does not strongly affect the circulation.  However, 
the metallicity controls the pressure of the photosphere (Spiegel et al. 2010, Lewis et al. 2010), and, since the
radiative time constant decreases with pressure, the temperature variation across the planet's disk (Lewis et al. 2010).  

We've also explored the heat distribution in GJ436b's atmosphere.  Instead of considering compositional effects, we use a
Newtonian relaxation scheme applied to the heat equation, which effectively creates a photosphere at the pressure levels 
most equivalent to those in Lewis et al.'s solar metallicity models .  
As a result, we derive temperature fields that resemble those
of Lewis' solar atmospheres.  In our GCM calculation 
the three-dimensional (3D) temperature distribution on GJ436b was simulated with the global atmospheric circulation model, BOB ("Built on Beowolf").  BOB solves the full primitive equations using a highly accurate and well-tested pseudospectral algorithm similar to that used in Thrastarson and Cho (2010).  The full details of the model can be found in Scott and Polvani (2008).  A series of simulations is performed, and a typical distribution from the series is presented in  Fig.~\ref{fig:Uat}.  In the simulation shown in the figures, the characteristic thermal relaxation time is dynamically adjusted in accordance with the planet's orbit, consistent with the used Newtonian cooling approximation (Salby 1996); the equilibrium temperature distribution is barotropic (vertically uniform) and zonally-symmetric (east-west symmetric), the latter in loose analogy with Venus.  The horizontal resolution of the simulation corresponds to T42 in each of the 20 layers.  For smooth fields, this is equivalent to at least 420$\times$210 grid resolution in each layer, compared to that in a conventional algorithm (e.g., Canuto et al. 1988).

Figure~\ref{fig:Uat} shows zonally-averaged zonal (eastward) wind distribution after 100~days 
(planet rotations) of integration, roughly 10~thermal relaxation times).  Figure~\ref{fig:Uat}
also shows the corresponding temperature distribution.  The basic flow structure is 
that of two strong, high latitude eastward jets and one weaker westward equatorial 
jet at the ``upper tropospheric'' ($\sim$300~mb) level.  Note that the jets are 
strongly barotropic, with only one modal variation in the vertical direction.  The jets 
are slowly strengthening and sharpening over time, which could be of dynamical significance.  This will be investigated in a future work.  Overall, the flow behavior is reflected by the temperature field (Figure~\ref{fig:Uat}), the 3D structure is fairly simple and in thermal wind balance (Salby 1996): although the temperature ranges are different, the basic shape looks very similar to that of the Earth's troposphere and lower-stratosphere.  The temperature decreases with height at all latitudes and then hints at a tropopause at high altitudes.  Dynamically generated temperature inversions are typically not present, or only very weakly present, in these set of calculations.  However, the calculations do not include several well known mechanisms for producing inversions, which may also affect the flow dynamics; these include absorbers at upper levels (Hubeny et al. 2003, Spiegel et al. 2010) and saturation of upwardly propagating waves (Watkins and Cho 2010).  Hence, any structures presented should be considered as requiring further confirmation by observations.

\begin{figure}
\begin{center}
\mbox { \includegraphics[width=8 cm]{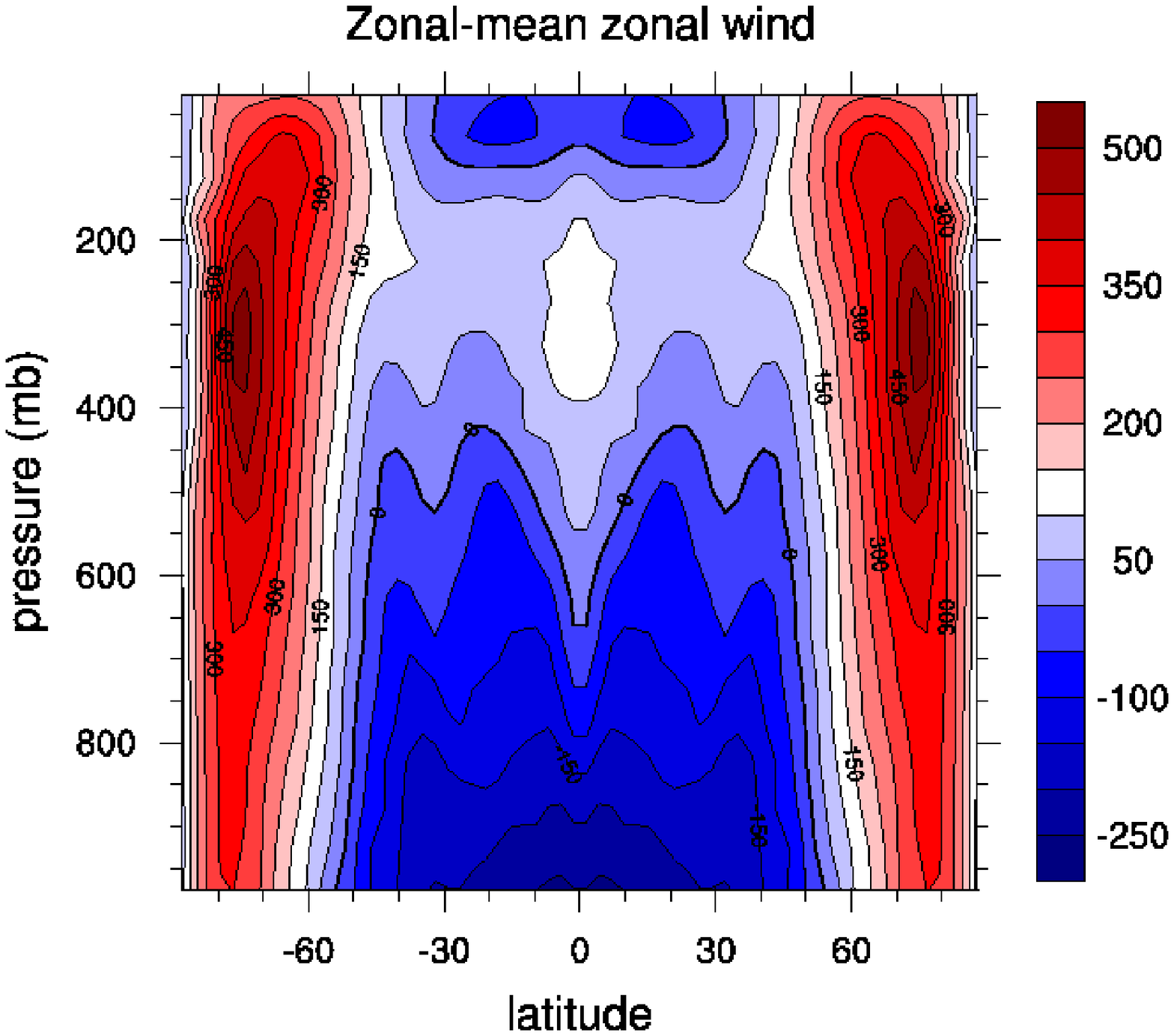}
\includegraphics[width=8 cm]{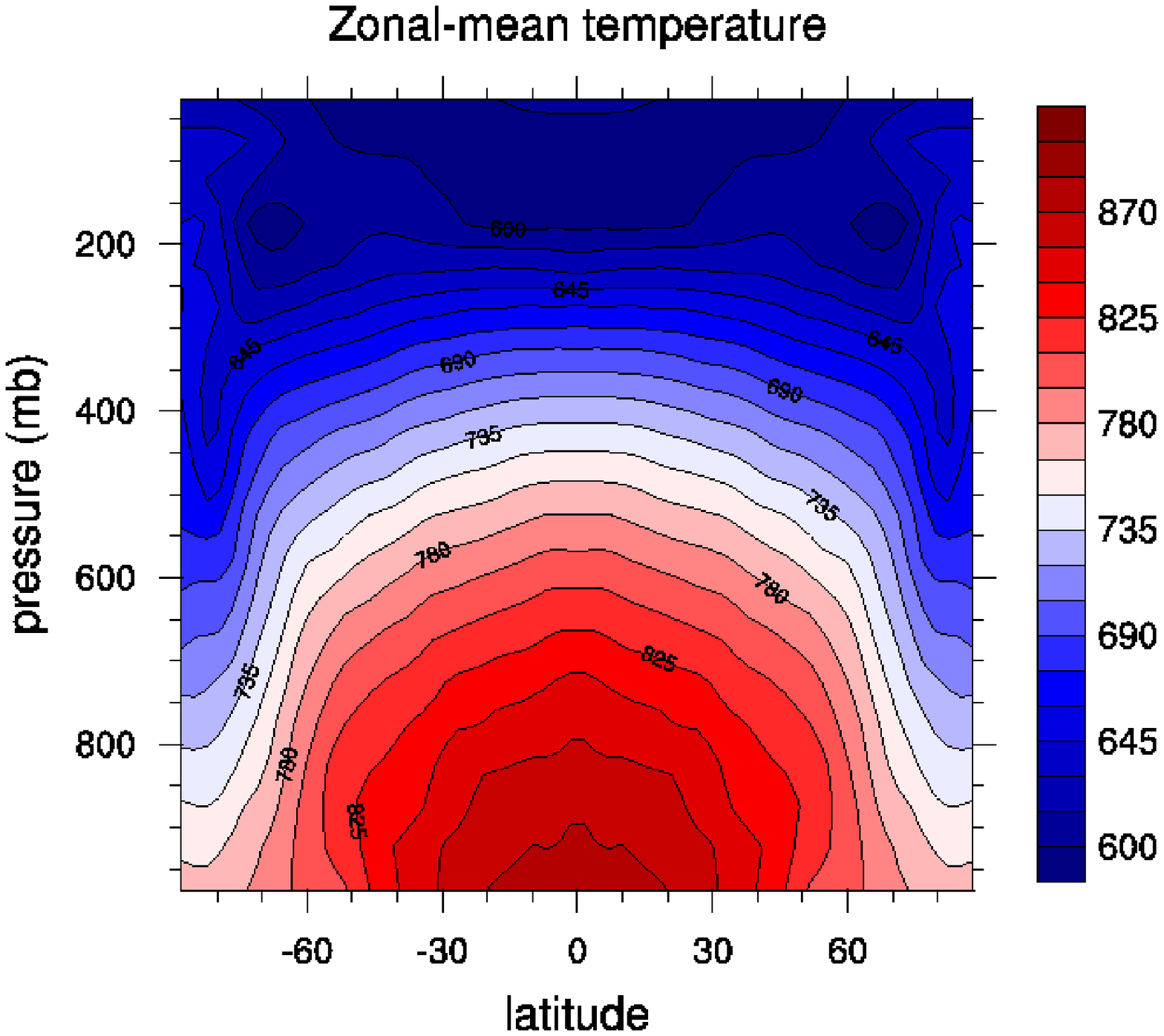}  }
\caption{Left: Zonally-averaged zonal (eastward) wind distribution.  In these
preliminary calculations, the distribution is characterised by three 
jets of roughly equal width in latitude.  Two jets are eastward, peaking 
at 400~m~s$^{-1}$ at $\sim$300~mb level and located at 70 degrees latitude 
in each hemisphere.   One jet is westward and located at the equator; the 
peaks of this jet,  maximal at the bottom (-250~m~s$^{-1}$), 
are off the equator at about 10 to 20~degrees latitude in each hemisphere.  
The overall structure is only weakly varying in the vertical direction 
and is strongly in thermal wind balance with the corresponding 
temperature structure.
\newline
Right: Zonally-averaged temperature distribution.  The distribution
is consistent with thermal wind balance  and upwelling
at the equator and downwelling near the equatorward flanks of the high
latitude jets.  The temperature simply decreases toward the poles and 
with height.  No inversion is present in this case. In few runs
a very weak inversion is sometimes observed.  However, several well
known mechanisms for producing inversions are not included in these
calculations.} 
\label{fig:Uat}
\end{center}
\end{figure}

}

\subsection{Modelling the transmission spectra of GJ436b}

To interpret the data we simulated transmission spectra of GJ436b, which account for the effects of molecular absorption (Tinetti et al. 2007a,b). { Our line by line radiative transfer model includes opacities due to $H_2O$, $CH_4$,  $NH_3$, $CO$ and $CO_2$. In our calculations here, we do not explore the range of molecular and temperature profiles that fit the data,
for lack of extensive wavelength coverage and spectral resolution. } We adopted  temperature profiles consistent with the calculations described in section 3.3 and with the secondary transit  observations by S2010. 
Using the BT2 line list for water (Barber et al. 2006) and the recently calculated list for ammonia (Yurchenko et al. 2010), we generated  the molecular opacities  at the appropriate temperatures. 
The complete line list for ammonia contains over two billion transitions, and is the most complete and accurate source of NH$_3$ opacity data. For the present study transitions involving rotationally excited states up to $J=23$ were explicitly considered.
Unfortunately line lists of methane at high temperature covering the spectral range $0.5-9~\mu$m are not yet available.  We  combined
HITRAN 2008 (Rothman et al., 2009), PNLL, Karkoschka and Tomasko (2010), and the high temperature measurements from Nassar and Bernath (2003) and Thi\'evin et al. (2008). 
For CO$_2$ we used HITEMP (Rothman et al.,
2010) and CDSD-1000 (Tashkun et al., 2003), for CO HITEMP was also used.  The contribution of H$_2$-H$_2$ at high
temperatures was taken from Borysow et al. (2001). The  opacity was interpolated to the temperature of each atmospheric layer. As collision induced absorption scales with the square of the pressure, the H$_2$-H$_2$ contribution becomes important for pressures higher than  1 bar. Given the plausible temperature range,  Na and K could be present in the atmosphere of GJ436b. Their opacities were estimated from Allard et al. (2003). 
 
\begin{figure}
\begin{center}
\includegraphics[width=14 cm]{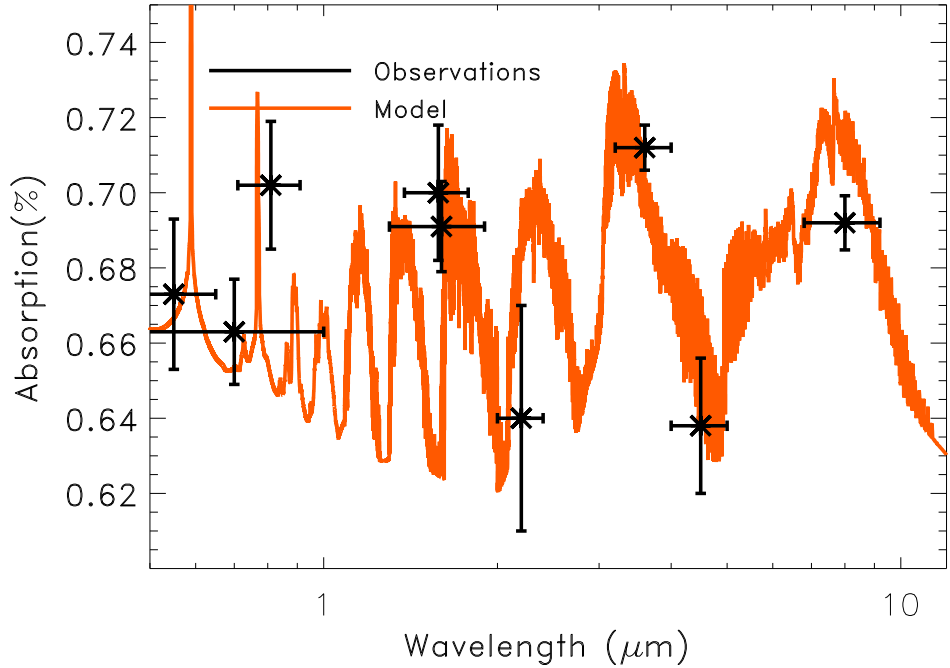}
\caption{\emph{
Simulated transmission spectrum of the transiting Hot Neptune GJ436b including the contribution of $CH_4$, $H_2$ and alkali metals in the wavelength range 0.5-9 $\mu$m. 
The simulation is a good fit to  the observations with Spitzer-IRAC at 3.6, 4.5 and 8 $\mu$m,
together with data collected by EPOXI in the range $0.5-1.0~\mu$m (Ballard et al., 2010), HST NICMOS  
in the range 1.2-1.8 $\mu$m (Pont et al., 2009), ground-based H-band (Alonso et al., 2008) and K-band (C\`aceres et al., 2009).  }} \label{fig:figmod}
\end{center}
\end{figure}

\begin{figure}
\begin{center}
\mbox{\includegraphics[width=8 cm]{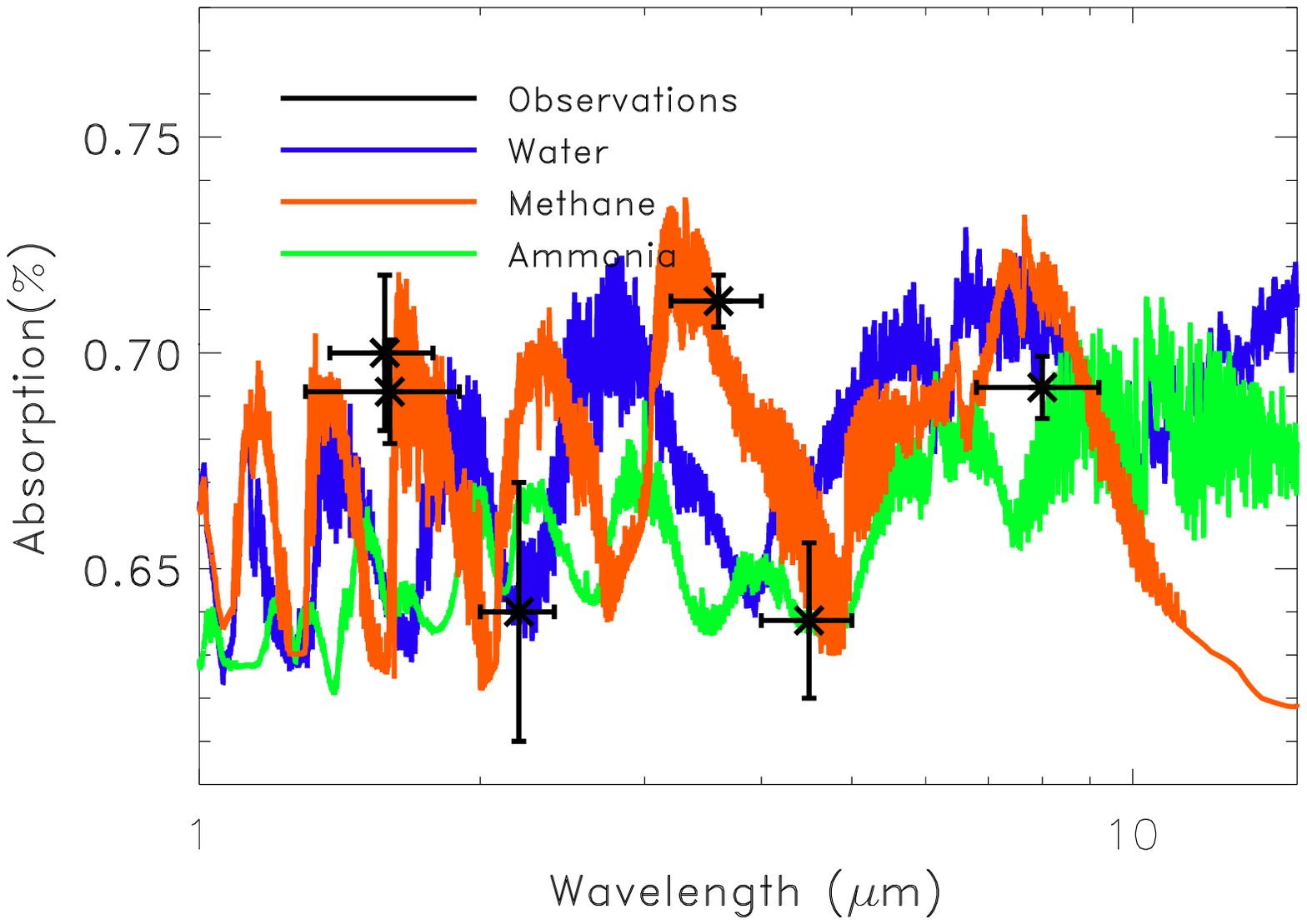}
\includegraphics[width=8 cm]{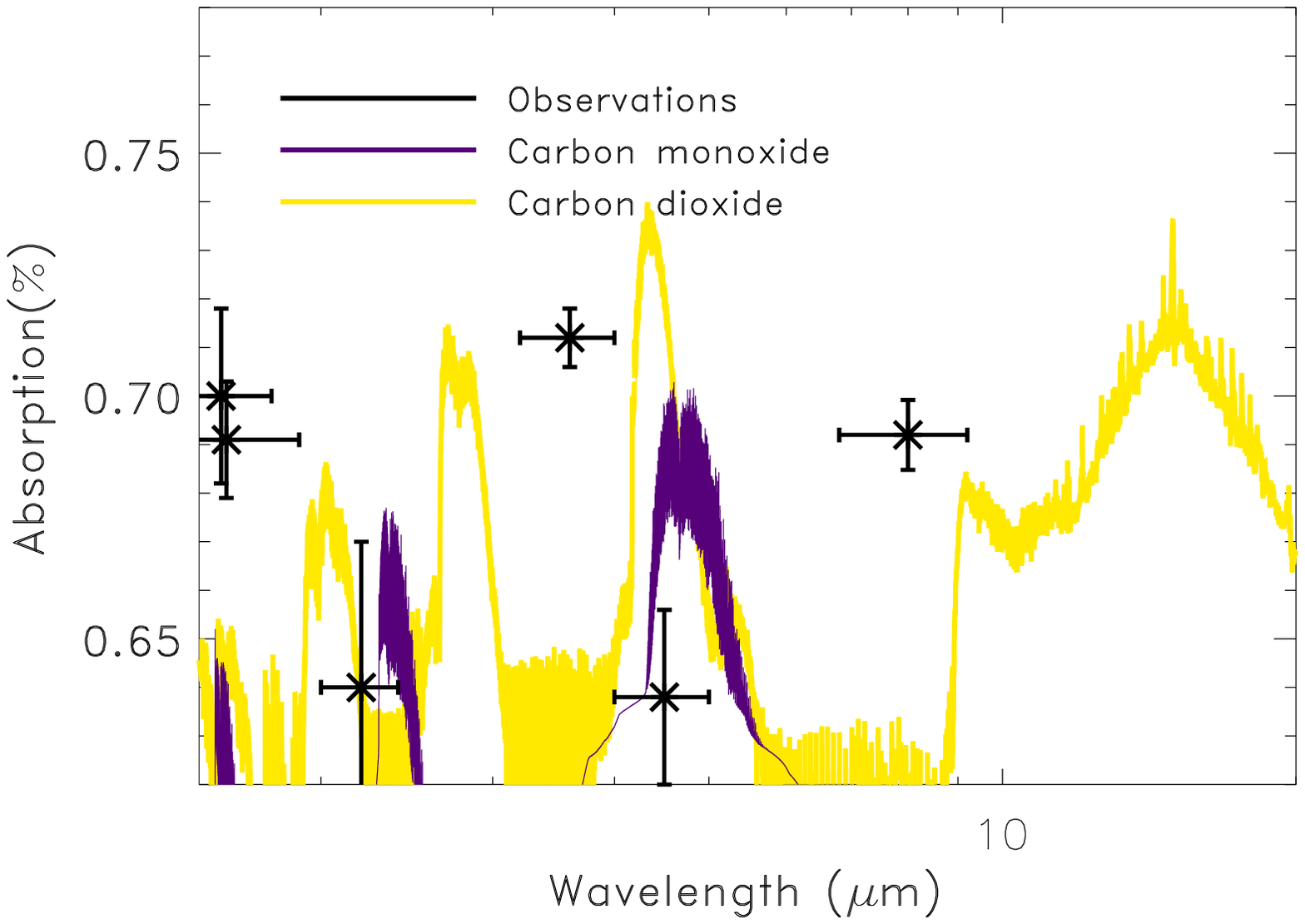} }
\caption{\emph{Simulated transmission spectra of the transiting Hot Neptune GJ436b. Left: we show here the contributions of methane, water, and ammonia. The relative modulation of the photometric bands can be explained mainly by methane.
Right: transmission spectra with the contribution  of carbon monoxide and dioxide. Compared to the observations, the signatures of these molecules show inconsistent behaviour.
   }} \label{fig:figmod2}
\end{center}
\end{figure}

Figure \ref{fig:figmod} shows the primary transit data of GJ436b, compared to a model spectrum for an atmosphere that
contains mainly  H$_2$ and methane, with a mixing ratio of $\sim 5 \cdot 10^{-4}$.  In the visible region,  our simulated spectrum is dominated by Rayleigh scattering and emission due to alkali metals. 
While we do not explore the range of solutions allowed by the
degeneracies in radius, temperature, and composition, in figure  \ref{fig:figmod2}, we can appreciate the contributions of other molecules. While water vapour and ammonia could potentially, but not crucially, be present,
the spectral features of CO and CO$_2$ display spectral patterns that oppose the transmission data, suggesting very limited, if any, abundances.

\begin{figure}
\begin{center}
\includegraphics[width=11 cm]{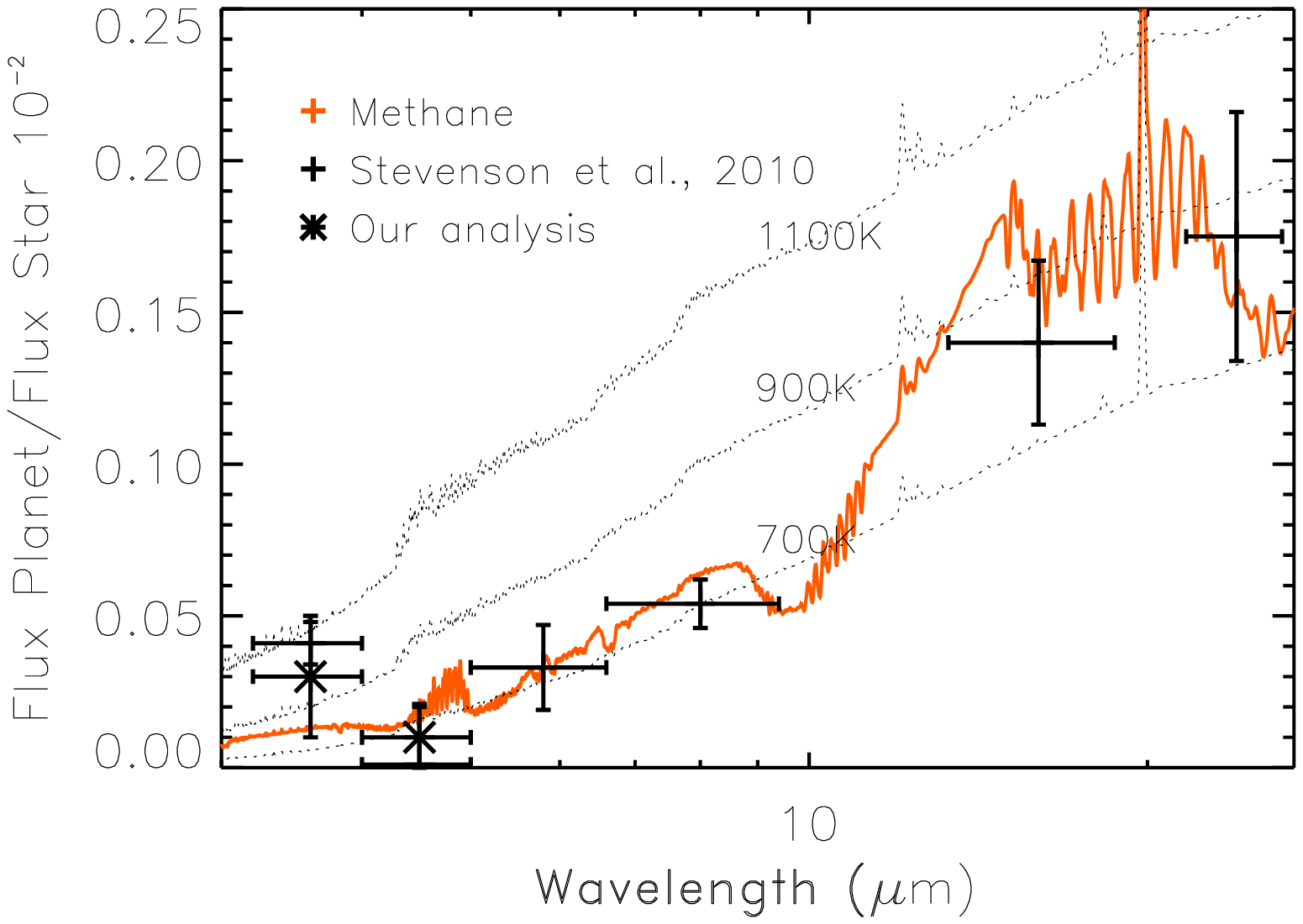}
\caption{\emph{Secondary transit observations analysed by S2010. Our analysis  suggests  larger error bars at 3.6 and 4.5 $\mu$m (starred points).
In colour, a simulated emission spectrum  with methane being predominant and a $T-P$ profile with an inversion at $\sim 10^{-2}$ bar. This solution is consistent with the interpretation of the primary transit data.
Dashed lines represent  black-body curves divided by stellar model synthetic spectra. 
 }} \label{fig:436bsec}
\end{center}
\end{figure}

We show in figure \ref{fig:436bsec} the secondary eclipse measurements published by S2010, along  
with one of our spectral models at 3.6 and 4.5 $~\mu$m { ( $0.03 \pm 0.02\%$\ and $0.01 \pm 0.01 \%$\ respectively, see appendix C for details). The interpretation of this data leads to a broad range highly degenerate solutions for the 
molecular composition and $T-P$ profile.  While a full study of the solution space is beyond the
scope of this paper, we find that a combination of molecular hydrogen and methane (as suggested by transmission data) with a $T-P$ profile containing an inversion at 10$^{-2}$-10$^{-3}$ bar fit the data quite well.  Such a methane-rich atmosphere is excluded by S2010 
analysis of the same data, which heavily relies the errors derived for two measurements at 3.6 and 4.5 $~\mu$m. Their errors bars 
prohibit the methane-rich atmospheres, which are allowed in our study, simply because we derive higher error bars than S2010. 
We find that both primary and secondary transit measurements are consistent with a CH$_4$-rich atmosphere, in agreement with thermochemical equilibrium models of exoplanet. Thus the disequilibria processes proposed by S2010 and Madhusudhan and Seager (2010) are not necessary to explain the data. However, still very little is known about the thermal and compositional profiles that characterize GJ436b.  Additional spectroscopic data are needed to break the degeneracy in composition and thermal profiles and distributions so that we can begin to understand the chemistry and dynamics of this Neptunian-sized planet. 

Shabram, Fortney, Green and Freedman (2010) compare and discuss the transit spectra for GJ436b presented here using chemical species and abundances adapted from Zahnle et al. (2009a) and Stevenson et al. (2010). In this paper and in a previous one (Fortney et al., 2010) they cast doubts on the validity of Tinetti's transmission models used to fit transit data for other hot-Jupiters (e.g. Tinetti et al., 2007; Swain et al., 2008, Beaulieu et al, 2009) on the basis they cannot reproduce the same results with their model. While the cause of the disagreement cannot be identified without a complete knowledge of all the parameters used as input to their model (e.g. gravity field, radius at the 1 bar level etc.), we notice here that Tinetti and Griffith obtain overlapping results when running their radiative transfer models in parallel, and are in agreement with Madhusudhan and Seager (2009) results (see e.g. fig. 6). Tinetti’s radiative transfer model uses the equations and observational geometry described in Seager and Sasselov (2000) and Brown (2001). 

Shabram et al. (2010) used an analytical relation between absorption cross-section and transit radius proposed by Lecavelier des Etangs et al. (2008) to validate their model. This analytical expression hypothesizes opacity cross-sections that vary as a power law: while this is correct in the case of a pure Rayleigh scattering atmosphere, the relation is unphysical when we consider molecular absorption, such as in the case of water vapour, as explained in classical radiative transfer books (e.g. Goody and Yung, 1961; Liou, 2002).

 } 

\section{Conclusions}
We have presented an analysis of seven primary transit observations of the hot Neptune GJ436b obtained with the Spitzer-IRAC camera.
  The final transit depth measurements obtained are $0.712 \pm 0.006\%$, $0.638 \pm 0.018 \%$, and $0.692 \pm 0.007 \%$ 
 respectively, at $3.6~\mu$m, $4.5~\mu$m, and $8~\mu$m. These new data taken together with the EPOXI, HST and ground-based $V, I, H$ and $K_s$  observations strongly suggests that methane is the dominant absorption in GJ436b's atmosphere. 
{ We point out that secondary eclipse data which samples the dayside atmosphere, are consistent with both methane abundant atmospheres as well as methane depleted atmosphere, proposed by Stevenson
et al., if the error bars are allowed to be larger at $3.6~\mu$m, $4.5~\mu$m, as proposed here.

In conclusion, transmission photometry of GJ436b indicates methane as the most carbon abundant 
species in the atmosphere at the terminator, while the molecular form of carbon on the day side is unclear; both methane rich and carbon monoxide rich atmospheres fit the data. Additional spectroscopic measurements are needed.}

\ack

  G.T. is supported by a Royal Society University Research Fellowship, D.K. and I. W. by STFC, R.J.B. by the Leverhulme Trust,
J.Y-K.C. is supported by the STFC PP/E001858/1 grant. IR acknowledges support from the Spanish Ministerio de Ciencia
e Innovaci\'on via grant AYA2009-06934.
We acknowledge the support by ANR-06-BLAN-0416 and the ``Programme Origine des Plan\`etes et de la Vie''.  This work is based on observations made with the \emph{Spitzer Space Telescopes}\ which is operated by the Jet Propulsion Laboratory, California Institute of Technology under a contract with NASA.





\appendix

\section{Pixel Phase Corrections for 3.6~$\mu$m and 4.5~$\mu$m}
\begin{figure*}
\begin{center}
\mbox{\includegraphics[angle=0,width=8. cm]{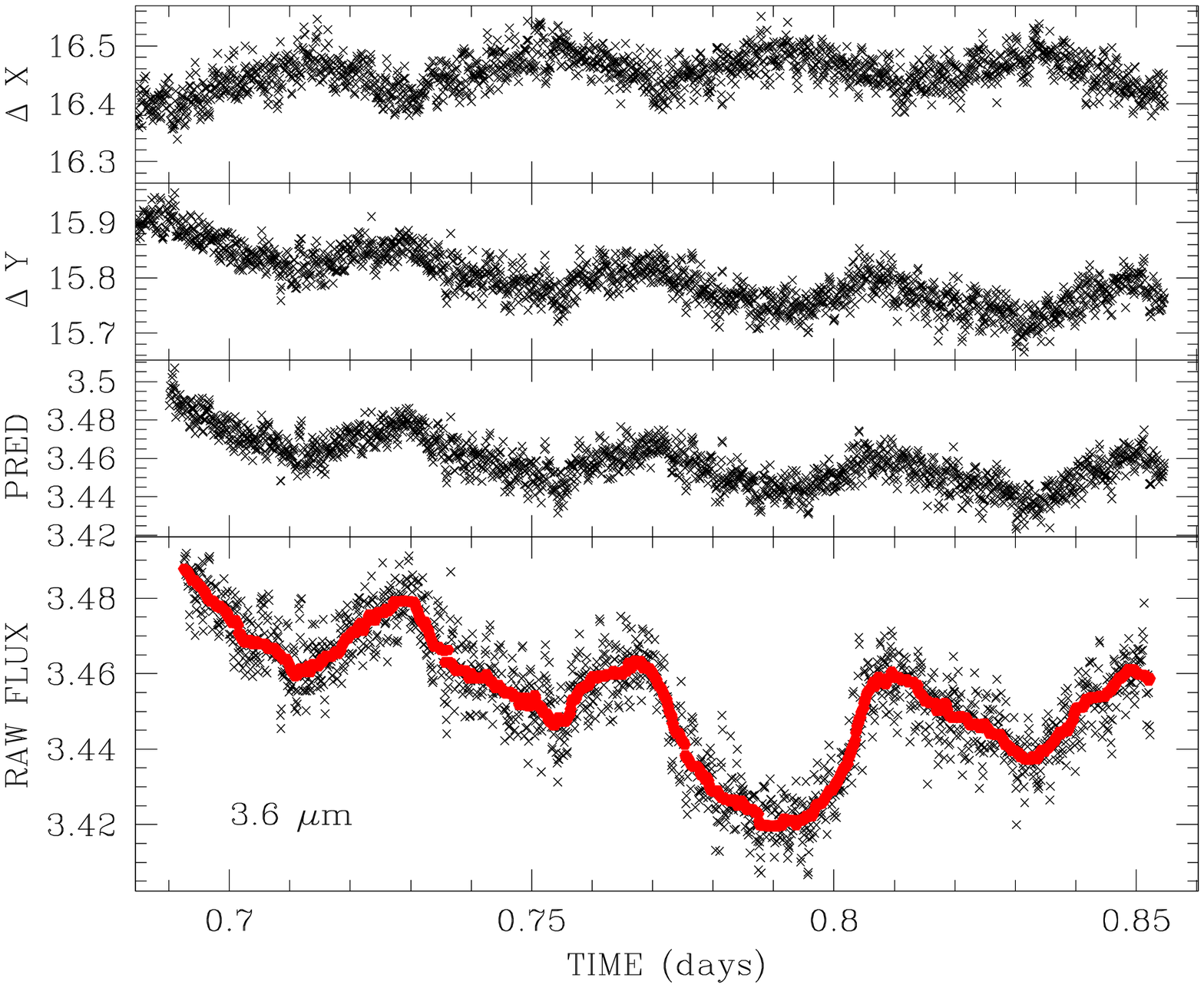} \includegraphics[angle=0,width=8. cm]{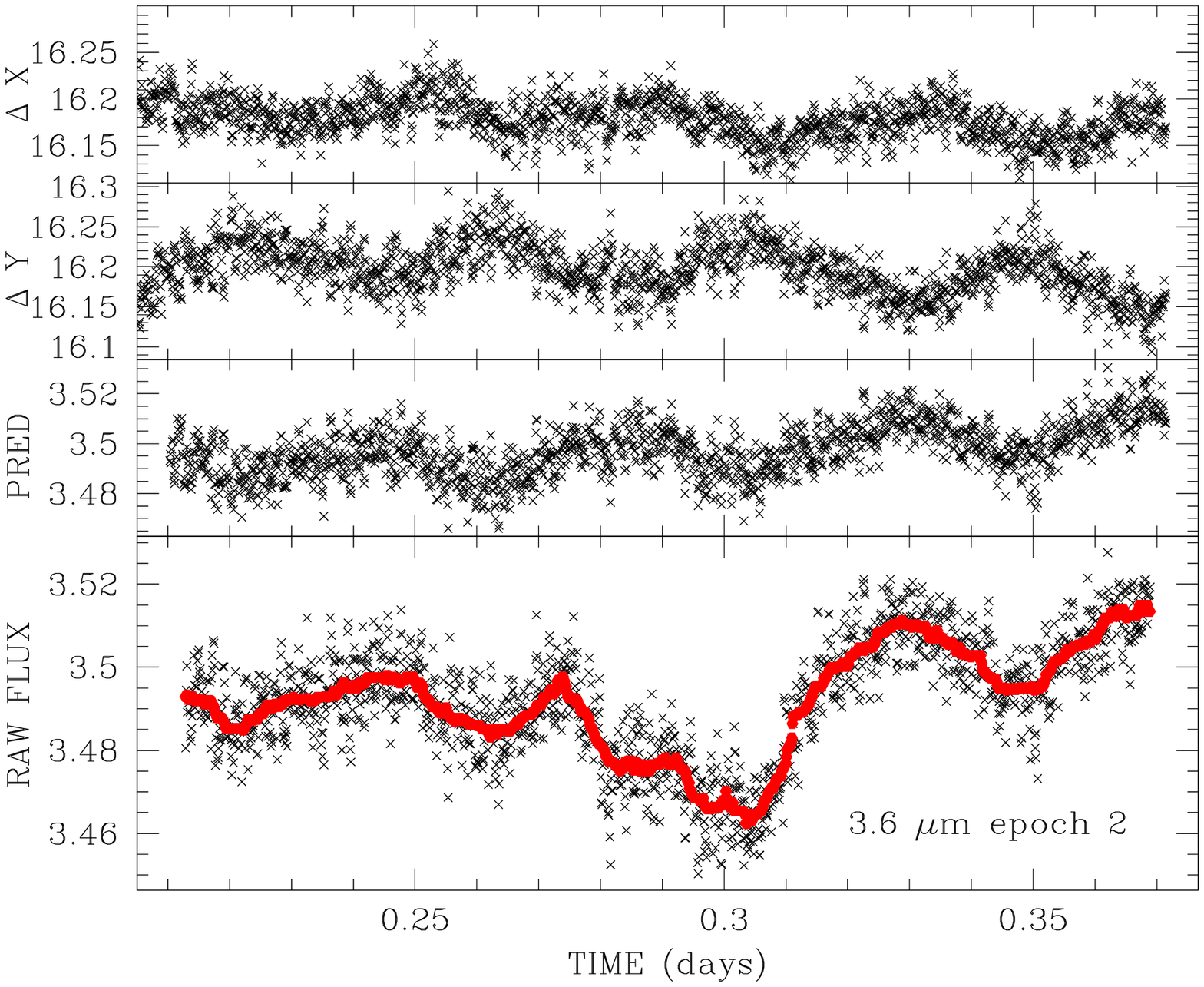} }
\mbox{\includegraphics[angle=0,width=8. cm]{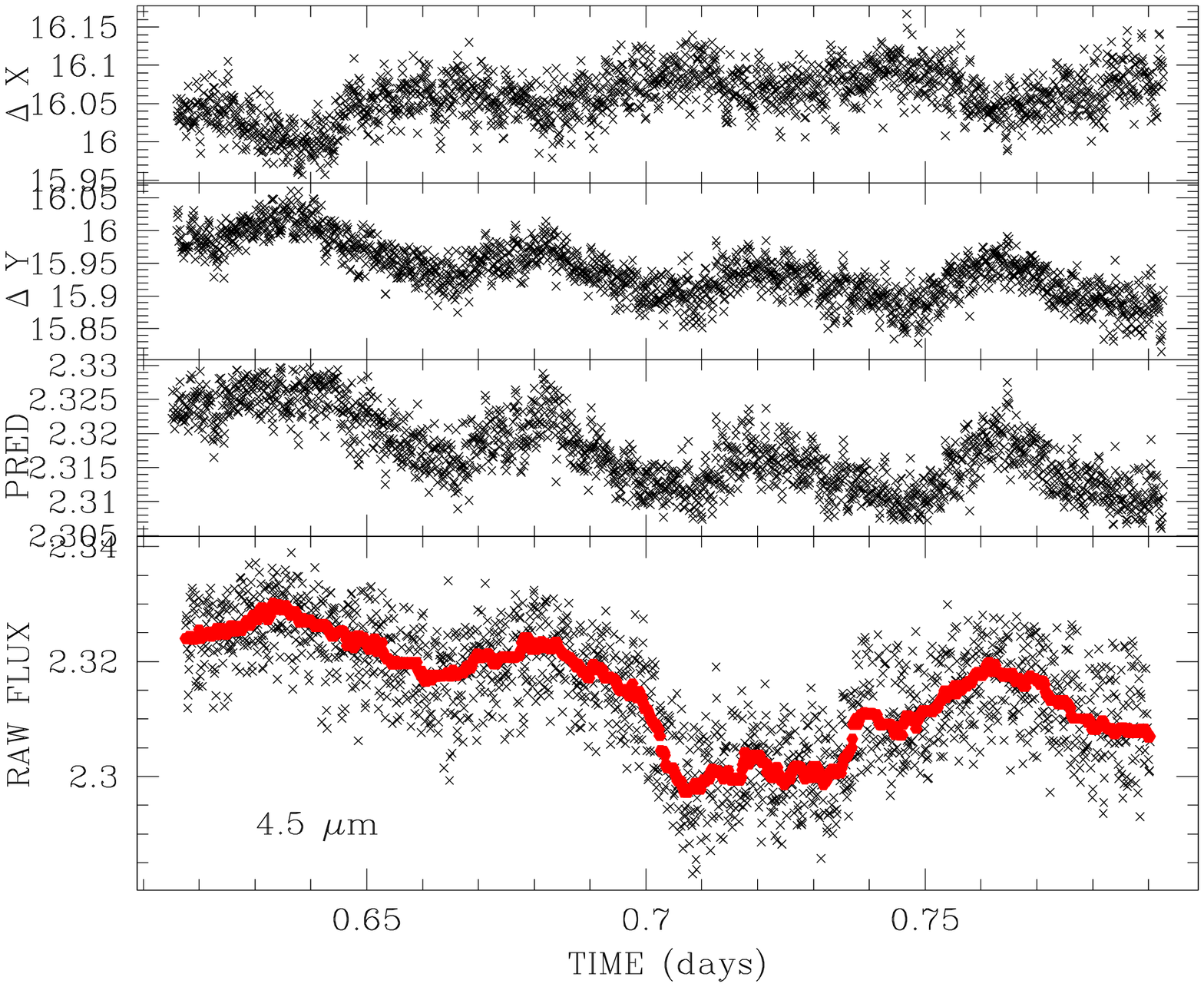} \includegraphics[angle=0,width=8. cm,]{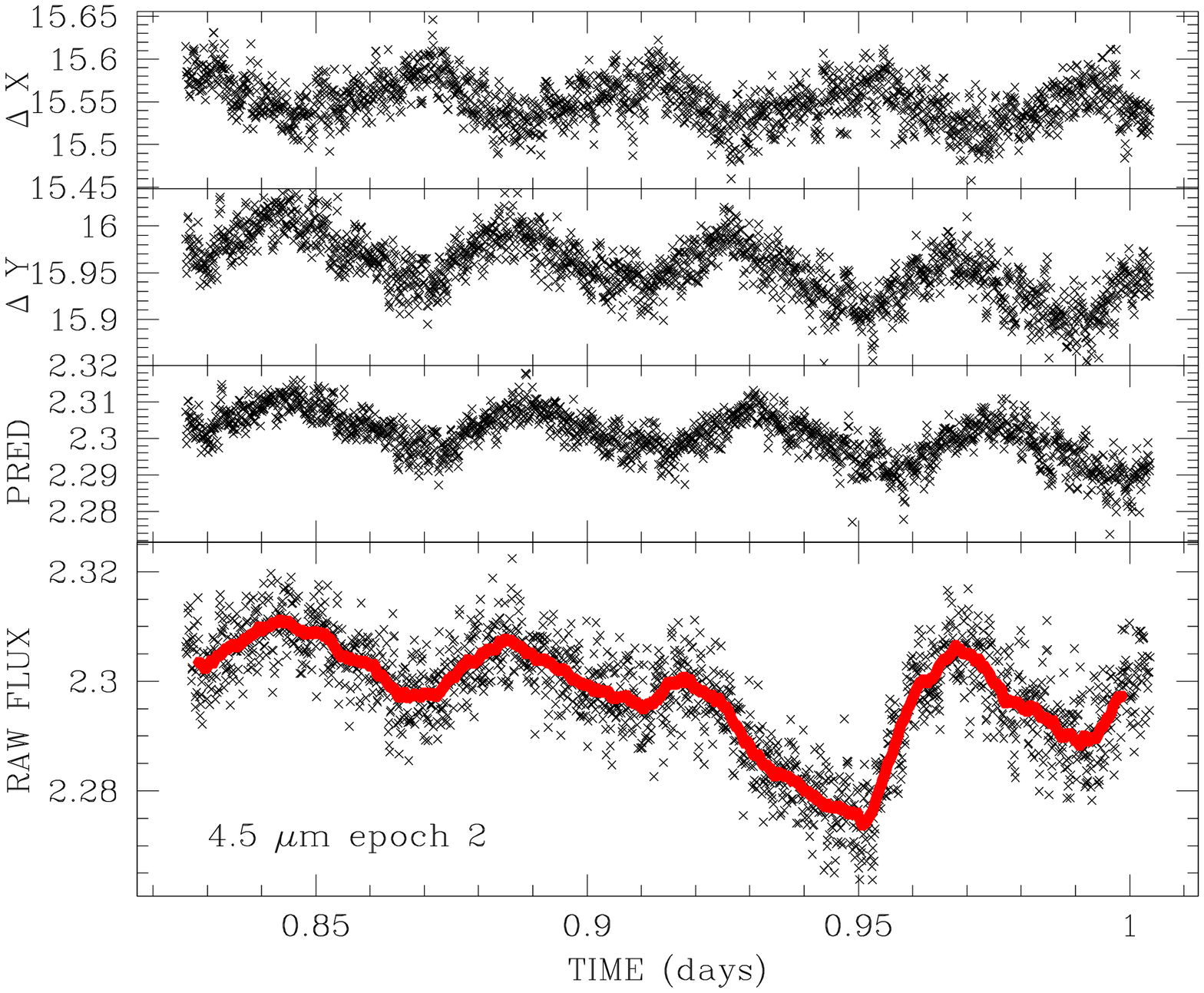} }
\caption{ Raw photometric data for 3.6 and 4.5  $\mu$m obtained with IRAC. Each sub-panel has the same structure showing from top to bottom: the variation of the centroid position in $X$, in $Y$, and lastly the 
predicted baseline flux using pixel phase correction. The lowest panel of each plot is the primary transit, 
and over-plotted the 50-point median-stack smoothing. They provide a synoptic
view of the systematic trends present in IRAC primary transit data.} \label{fig:lc3.6-4.5}
\end{center}
\end{figure*}

The raw data at 3.6~$\mu$m and 4.5~$\mu$m are shown in  Figure~\ref{fig:lc3.6-4.5}.

\subsection{Four families of pixel phase corrections}

Pixel phase variations can be corrected for by correlating the observed flux to 
the pixel position. Regardless of what co-ordinate system one wishes to adopt, 
two parameters are required to describe the position of the centroid within the 
pixel. In addition, there is a possibility of a temporal dependence, both in the
absolute flux and/or the intrapixel response function itself. This therefore 
extends the number of parameters required to fully describe any correlations to
three. In this work, we consider four families of pixel phase correction:

\begin{itemize}
\item \textbf{CN} - Cartesian co-ordinate system for spatial parameters, without time components: $x, y$
\item \textbf{CT} - Cartesian co-ordinate system for spatial parameters, with time components: $x, y, t$
\item \textbf{PN} - Polar co-ordinate system for spatial parameters, without time components: $r, \theta$
\item \textbf{PT} - Polar co-ordinate system for spatial parameters, with time components: $r,\theta, t$
\end{itemize}

\subsection{Model exploration}

For any one given family, there is a wide range of possible combinations of the 
parameters to fit for. In this work, we will consider symmetric polynomial 
expansions of the dependent parameters. For each family, we have tried linear 
($n=1$), quadratic ($n=2$), cubic ($n=3$) and quartic ($n=4$) symmetric 
polynomial expansions for the form of the pixel phase model.

For the non-temporal families, the number of degrees of freedom 
in the fit will follow the triangular number set (3,6,10,15...). In contrast, 
the temporal-families will follow the tetrahedral number set (4,10,20,35,...).

Below, we provide the explicit forms of all of these models.
For the CN family (Cartesian Non-temporal):

\begin{itemize}
\item[{$n=1$}] $a_1$ + $a_2 x$ + $a_3 y$ \\
\item[{$n=2$}] $a_1$ + $a_2 x$ + $a_3 y$ + $a_4 x^2$ + $a_5 y^2$ + $a_6 x y$ \\
\item[{$n=3$}] $a_1$ + $a_2 x$ + $a_3 y$ + $a_4 x^2$ + $a_5 x y$ + $a_6 y^2$
              +$a_7 x^3$ + $a_8 x^2 y$ + $a_9 x y^2$ + $a_{10} y^3$ \\
\item[{$n=4$}] $a_1$ + $a_2 x$ + $a_3 y$ + $a_4 x^2$ + $a_5 x y$ + $a_6 y^2$
              +$a_7 x^3$ + $a_8 x^2 y$ + $a_9 x y^2$ + $a_{10} y^3$
              +$a_{11} x^4$ + $a_{12} x^3 y$ + $a_{13} x^2 y^2$ + $a_{14} x y^3$
              +$a_{15} y^4$
\end{itemize}

For the CT family (Cartesian Temporal), we used:

\begin{itemize}
\item[{$n=1$}] $b_1$ + $b_2 x$ + $b_3 y$ + $b_4 t$ \\
\item[{$n=2$}] $b_1$ + $b_2 x$ + $b_3 y$ + $b_4 t$
              +$b_5 x^2$ + $b_6 x y$ + $b_7 x t$ + $b_8 y^2$ + $b_9 y t$ + $b_{10} t^2$ \\
\item[{$n=3$}] $b_1$ + $b_2 x$ + $b_3 y$ + $b_4 t$
              +$b_5 x^2$ + $b_6 x y$ + $b_7 x t$ + $b_8 y^2$ + $b_9 y t$ + $b_{10} t^2$
              +$b_{11} x^3$ + $b_{12} x^2 y$ + $b_{13} x^2 t$ + $b_{14} x y^2$ + $b_{15} x y t$ + $b_{16} x t^2$
              +$b_{17} y^3$ + $b_{18} y^2 t$ + $b_{19} y t^2$ + $b_{20} t^3$ \\
\item[{$n=4$}] $b_1$ + $b_2 x$ + $b_3 y$ + $b_4 t$
              +$b_5 x^2$ + $b_6 x y$ + $b_7 x t$ + $b_8 y^2$ + $b_9 y t$ + $b_{10} t^2$
              +$b_{11} x^3$ + $b_{12} x^2 y$ + $b_{13} x^2 t$ + $b_{14} x y^2$ + $b_{15} x y t$ + $b_{16} x t^2$
              +$b_{17} y^3$ + $b_{18} y^2 t$ + $b_{19} y t^2$ + $b_{20} t^3$
              +$b_{21} x^4$ + $b_{22} x^3 y$ + $b_{23} x^3 t$ 
              +$b_{24} x^2 y^2$ + $b_{25} x^2 y t$ + $b_{26} x^2 t^2$
              +$b_{27} x y^3$ + $b_{28} x y^2 t$ + $b_{29} x y t^2$ + $b_{30} x t^3$
              +$b_{31} y^4$ + $b_{32} y^3 t$ + $b_{33} y^2 t^2$ + $b_{34} y t^3$ + $b_{35} t^4$
\end{itemize}

For the PN family, we have:

\begin{itemize}
\item[{$n=1$}] $c_1$ + $c_2 r$ + $c_3 \theta$ \\
\item[{$n=2$}] $c_1$ + $c_2 r$ + $c_3 \theta$ + $c_4 r^2$ + $c_5 \theta^2$ + $c_6 r \theta$ \\
\item[{$n=3$}] $c_1$ + $c_2 r$ + $c_3 \theta$ + $c_4 r^2$ + $c_5 r \theta$ + $c_6 \theta^2$
              +$c_7 r^3$ + $c_8 r^2 \theta$ + $c_9 r \theta^2$ + $c_{10} \theta^3$ \\
\item[{$n=4$}] $c_1$ + $c_2 r$ + $c_3 \theta$ + $c_4 r^2$ + $c_5 r \theta$ + $c_6 \theta^2$
              +$c_7 r^3$ + $c_8 r^2 \theta$ + $c_9 r \theta^2$ + $c_{10} \theta^3$
              +$c_{11} r^4$ + $c_{12} r^3 \theta$ + $c_{13} r^2 \theta^2$ + $c_{14} r \theta^3$
              +$c_{15} \theta^4$
\end{itemize}

For the PT family (Polar Temporal), we used:

\begin{itemize}
\item[{$n=1$}] $d_1$ + $d_2 r$ + $d_3 \theta$ + $d_4 t$ \\
\item[{$n=2$}] $d_1$ + $d_2 r$ + $d_3 \theta$ + $d_4 t$
              +$d_5 r^2$ + $d_6 r \theta$ + $d_7 r t$ + $d_8 \theta^2$ + $d_9 \theta t$ + $d_{10} t^2$ \\
\item[{$n=3$}] $d_1$ + $d_2 r$ + $d_3 \theta$ + $d_4 t$
              +$d_5 r^2$ + $d_6 r \theta$ + $d_7 r t$ + $d_8 \theta^2$ + $d_9 \theta t$ + $d_{10} t^2$
              +$d_{11} r^3$ + $d_{12} r^2 \theta$ + $d_{13} r^2 t$ + $d_{14} r \theta^2$ + $d_{15} r \theta t$ + $d_{16} r t^2$
              +$d_{17} \theta^3$ + $d_{18} \theta^2 t$ + $d_{19} \theta t^2$ + $d_{20} t^3$ \\
\item[{$n=4$}] $d_1$ + $d_2 r$ + $d_3 \theta$ + $d_4 t$
              +$d_5 r^2$ + $d_6 r \theta$ + $d_7 r t$ + $d_8 \theta^2$ + $d_9 \theta t$ + $d_{10} t^2$
              +$d_{11} r^3$ + $d_{12} r^2 \theta$ + $d_{13} r^2 t$ + $d_{14} r \theta^2$ + $d_{15} r \theta t$ + $d_{16} r t^2$
              +$d_{17} \theta^3$ + $d_{18} \theta^2 t$ + $d_{19} \theta t^2$ + $d_{20} t^3$
              +$d_{21} r^4$ + $d_{22} r^3 \theta$ + $d_{23} r^3 t$ 
              +$d_{24} r^2 \theta^2$ + $d_{25} r^2 \theta t$ + $d_{26} r^2 t^2$
              +$d_{27} r \theta^3$ + $d_{28} r \theta^2 t$ + $d_{29} r \theta t^2$ + $d_{30} r t^3$
              +$d_{31} \theta^4$ + $d_{32} \theta^3 t$ + $d_{33} \theta^2 t^2$ + $d_{34} \theta t^3$ + $d_{35} t^4$
\end{itemize}

\subsection{Model selection}

As we increase the number of degrees of freedom, $k$, any fit we perform will 
naturally produce a lower merit function. The question therefore is at which 
point do we stop adding degrees of freedom? S2010 proposed using the Bayesian 
Information Criterion (BIC) to make this determination. BIC has the advantage 
of penalising models for being overly-complex and thus will not decrease 
ad-infinitum as $n$ increases. The optimum model is given by that which yields 
the lowest BIC. Defining $x_i$ as the residual of a fit, $\sigma_i$ as the 
measurement uncertainty and $N$ as the number of data points, the expression for
BIC is given by:

\begin{equation}
\mathrm{BIC} = \sum_{i=1}^N (x_i/\sigma_i)^2 + k \mathrm{ln}N
\end{equation}

In Table~\ref{table:fullbic} we provide the full list of all models tried to
correct the pixel phase effects for the four affected time series.

\begin{table*}
{\tiny
\begin{tabular}{l l l l l} 
\hline 
Order, $n$ & CN BIC & CT BIC & PN BIC & PT BIC  \\
\hline
\multicolumn{5}{c}{ {\it {\bf Channel 1}} }   \\
\hline
1 & 1300.171       & 1304.982       & 1310.678       & 1315.350       \\
2 & 1276.464       & {\bf 1284.769} & 1278.686       & {\bf 1286.245} \\
3 & {\bf 1275.677} & 1317.442       & {\bf 1272.918} & 1314.945       \\
4 & 1293.197       & 1394.872       & 1290.468       & 1390.206       \\
\hline
\multicolumn{5}{c}{ {\it {\bf Channel 2}} }  \\
\hline 
1 & {\bf 1444.680} & {\bf 1429.981} & {\bf 1444.380} & {\bf 1429.626} \\
2 & 1445.941       & 1446.977       & 1446.573       & 1447.901       \\
3 & 1450.939       & 1480.064       & 1450.999       & 1479.110       \\
4 & 1475.250       & 1577.162       & 1476.360       & 1577.638       \\
\hline  
 \multicolumn{5}{c}{ {\it {\bf Channel 5}} } \\ 
\hline  
1 & 1264.150       & {\bf 1196.149} & 1308.062       & 1245.248       \\
2 & 1266.709       & 1206.635      & 1268.543       & {\bf 1208.899} \\
3 & {\bf 1259.664} & 1246.510      & {\bf 1260.148} & 1249.250       \\
4 & 1274.027       & 1336.543      & 1273.779       & 1336.462       \\
\hline  
 \multicolumn{5}{c}{ {\it {\bf Channel 8}} } \\ 
\hline  
1 & 1552.284       & 1556.485       & 1526.502       & 1530.995       \\
2 & {\bf 1504.001} & {\bf 1484.829} & {\bf 1506.415} & {\bf 1492.758} \\
3 & 1526.905       & 1555.750       & 1526.351       & 1551.959       \\
4 & 1560.801       & 1652.530       & 1560.612       & 1652.909       \\
\hline 
\end{tabular} }
\caption{\emph{BIC values of all the models we tried to correct the pixel phase
effect. For each channel, 4 families of corrective procedure were attempted
(CN, CT, PN, PT; where C = Cartesian, P = Polar, N = non-temporal, T = temporal)
with 4 orders of complexity each ($n=1,2,3,4$). The lowest BIC for each family
is highlighted in bold.}} 
\label{table:fullbic} 
\end{table*}

\subsection{Final Model Values}

For each time series, we show the best-fit model parameters in 
Table~\ref{table:modelparams}. For all dimensions (i.e. $x$,$y$, $r$, $\theta$ 
and $t$), the median of the array is subtracted first before fitting. This step
serves to reduce interparameter correlations. For example, for $t$, which is in
BJD, the BJD value is much larger than the duration of the measurement and thus
fitting a function $\alpha + \beta t$ would give rise to a very large 
correlation between $\alpha$ and $\beta$ and thus very large errors. A better
approach is to move the pivot to the median of the time stamps which serves as
a far improved pivot point. The same it true for the other parameters as well.

\begin{table*}
{\tiny
\begin{tabular}{l l l} 
\hline 
Parameter & Best fit & Standard Error  \\
\hline
\multicolumn{3}{c}{ {\it {\bf Channel 1}} }\\
\hline
$c_1$ & 34590.6 & 5.8 \\
$c_2$ & 1670 & 290 \\
$c_3$ & -32.2 & 2.5 \\
$c_4$ & 11400 & 4600 \\
$c_5$ & -396 & 78 \\
$c_6$ & -0.64 & 0.76 \\
$c_7$ & -41000 & 21000 \\
$c_8$ & 0 & 630 \\
$c_9$ & 545 & 13 \\
$c_{10}$ & 5.79 & 0.25 \\
\hline
\multicolumn{3}{c}{ {\it {\bf Channel 2}} }\\
\hline 
$d_1$ & 23199.7 & 3.4 \\
$d_2$ & 30 & 61 \\
$d_3$ & -13.5 & 1.0 \\
$d_4$ & -760 & 31 \\
\hline  
 \multicolumn{3}{c}{ {\it {\bf Channel 5}} }\\ 
\hline  
$b_1$ & 34973.24 & 0.84 \\
$b_2$ & -757 & 40 \\
$b_3$ & -2963 & 28 \\
$b_4$ & 159 & 18 \\
\hline  
 \multicolumn{3}{c}{ {\it {\bf Channel 8}} }\\ 
\hline  
$b_1$ & 23003.3 & 1.9 \\
$b_2$ & -368 & 49 \\
$b_3$ & 1812 & 49 \\
$b_4$ & -112 & 39 \\
$b_5$ & 4700 & 1300 \\
$b_6$ & 4400 & 2000 \\
$b_7$ & -800 & 1200 \\
$b_8$ & -700 & 1200 \\
$b_9$ & 2900 & 1300 \\
$b_{10}$ & 3420 & 600 \\ 
\hline 
\end{tabular} }
\caption{\emph{Best-fitted model parameters for each time series, where we only
present the parameters of the final favoured model (determined using the 
Bayesian Information Criterion). Best-fit parameters computed using a
Levenberg-Marquardt algorithm. All errors quoted to 2 significant figures
and corresponding best-fit values to the equivalent number of decimal places.}} 
\label{table:modelparams} 
\end{table*}

\subsection{Fitting procedure}

The final point we wish to address is how these fits should be performed. There 
are currently two schools of thought. The first is to exclude the data where the
astrophysical signal is expected and use the baseline data as a calibration tool
(e.g. Knutson et al. 2007 ; Beaulieu et al. 2010). In this approach, we simply 
perform a least-squares fit of a given model to the flux counts.

The second approach is to fit for both the astrophysical signal and the 
systematic model simultaneously. An obvious advantage of this approach is that 
correlations between the transit depth and the selected free parameters can be 
investigated. However there are two concerns with this approach: the first one 
occurs when one is faced with an astrophysical signal of the same phase and 
time-scale as the systematics. For transiting planets like HD 209458b, 
HD 189733b or HD 80606 we have several cycles of the pixel phase effect within 
the transit. This, unfortunately, is not the case for transiting planets such 
as CoRoT-7b, or even worse GJ1214b and GJ436b, where the transit duration is 
similar to the pixel phase time-scale.

The second critical problem is  modelling  the astrophysical signal. If one 
assumes a simple transit or eclipse then there are no concerns. Unfortunately, a
simple transit model may be insufficient at the level of precision obtained by 
Spitzer. Particularly when observing M-dwarfs, one cannot exclude the 
possibility of starspots affecting the transit signal. A starspot crossing would
induce a bump in the transit shape which would bear a remarkable resemblance to 
half of a pixel-phase period. If such a bump existed and a simultaneous fit was
performed, the systematic correction would try to model the bump as part of the 
pixel phase response and thus lead to essentially an erroneous correction.

Other effects such as an exomoon transit (Simon et al. 2009), planetary 
oblateness (Seager \& Hui 2002), atmospheric lensing (Sidis \& Sari 2010) and
rings (Barnes \& Fortney 2004), to name a few, can all induce anomalous transit
features as well. In essence, by fitting for systematics plus a transit signal
across all of the data simulatenously, one has already chosen what one will
discover.

This subtle point has not been previously emphasised in the exoplanet 
literature, but it does raise severe concerns about any results found using such
methods, particularly for spotty stars. In the case of GJ436, the presence of 
star spots has already been reported by Demory et al. (2007) and more recently 
by Ballard et al. (2010) using EPOXI. We will return to the issue of GJ436's 
activity in the subsequent sections. In general, it is preferable to err on the 
side of caution: there therefore  exists an additional strong motivation to 
exclude the transit signal when attempting to apply systematic corrections. It 
is this approach that will be adopted for the remainder of our analysis.

\subsection{The unique problems of GJ436}

Whilst BIC offers a clean, statistical way of discriminating between the various
models, there should be some caution in uncritically using this tool, 
particularly for GJ436. GJ436b is  unique in that the duration of the transit is
60 mins and the pixel phase effect has a period between 50 and 60 mins. The 
close proximity of these two time-scales means that any single pixel phase 
correction, even one which exhibits the lowest BIC, should be taken with 
caution. The problem is exacerbated by the fact that the transit of GJ436b is 
relatively shallow ($\sim7$~mmag) and actually comparable to the pixel phase 
amplitude ($\sim5$~mmag). We also note that the secondary eclipse has a very 
similar duration to the primary transit due to the argument of periapse being 
close to $\pi/2$.

\begin{figure*}
\begin{center}
\mbox{\includegraphics[width=8 cm]{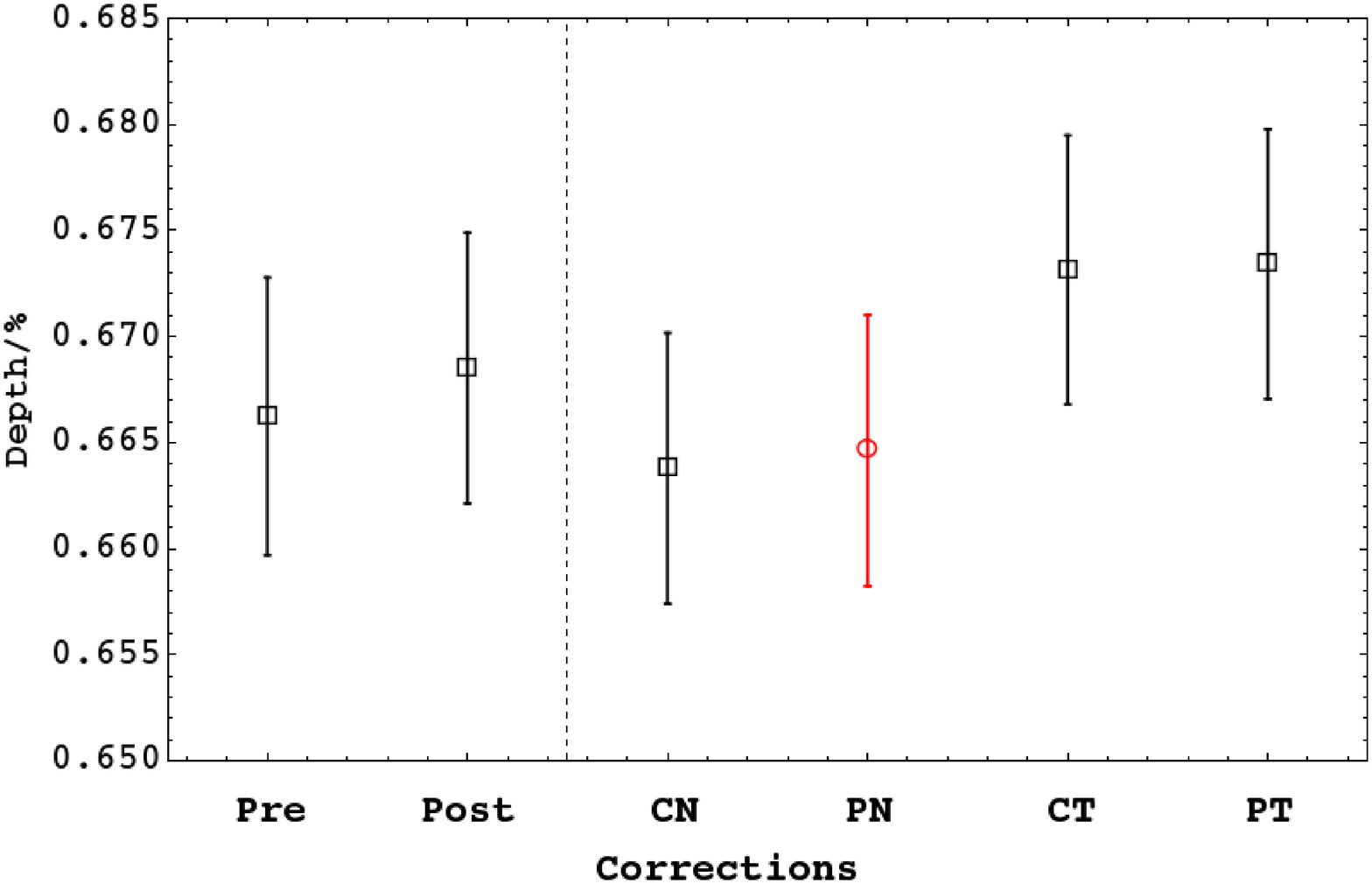}\includegraphics[width=8 cm]{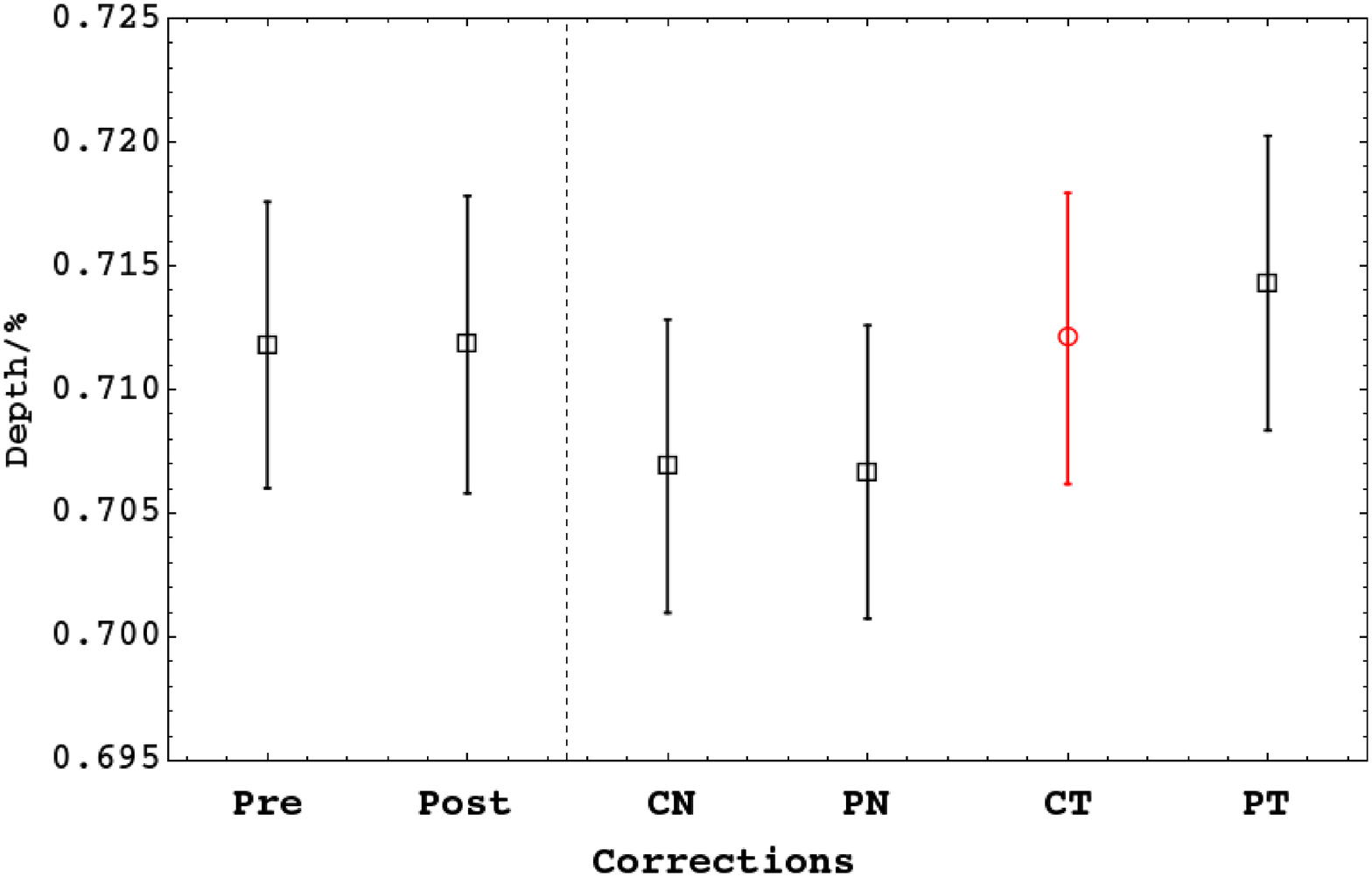} }
\mbox{\includegraphics[width=8 cm]{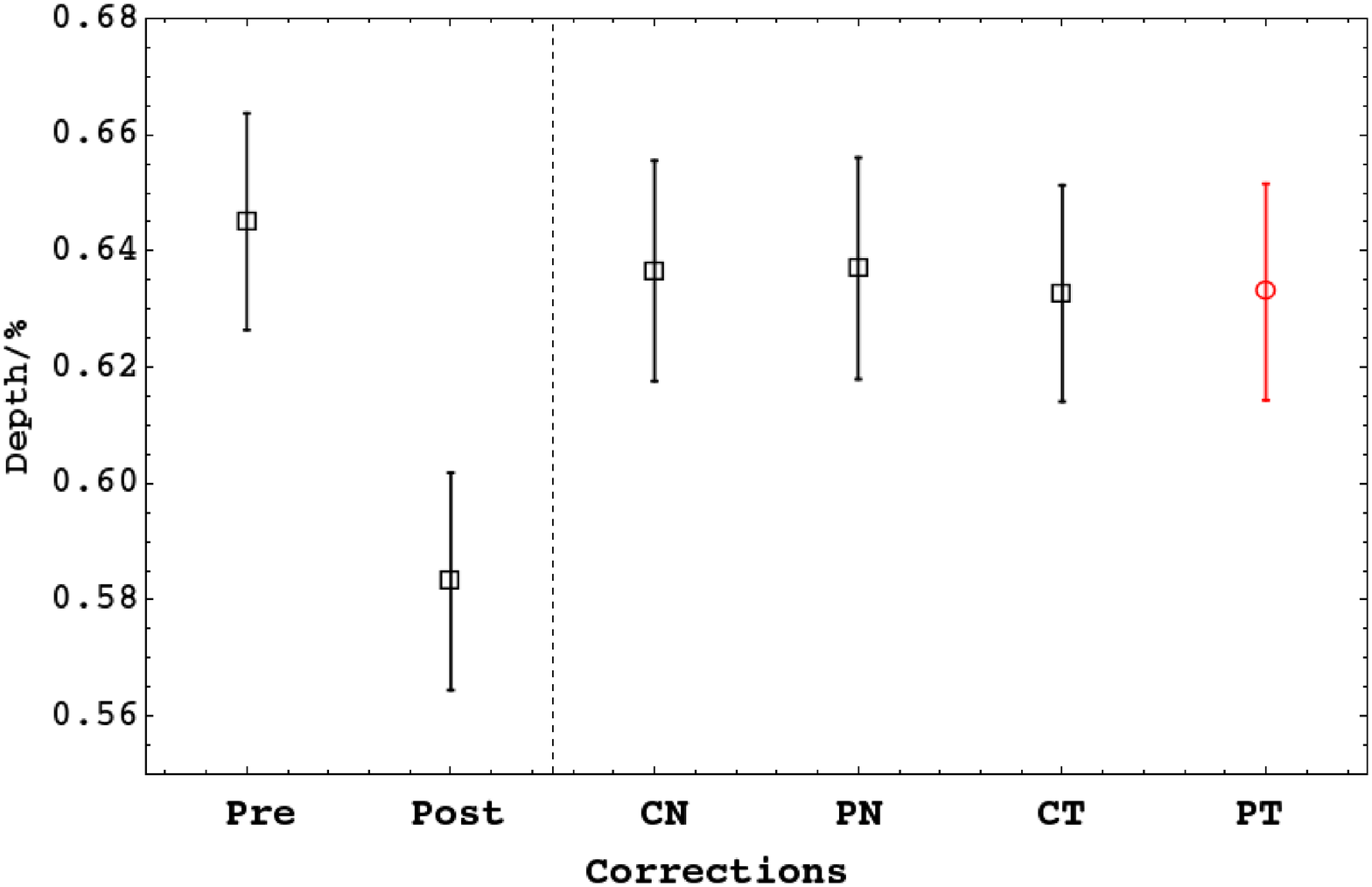}\includegraphics[width=8 cm]{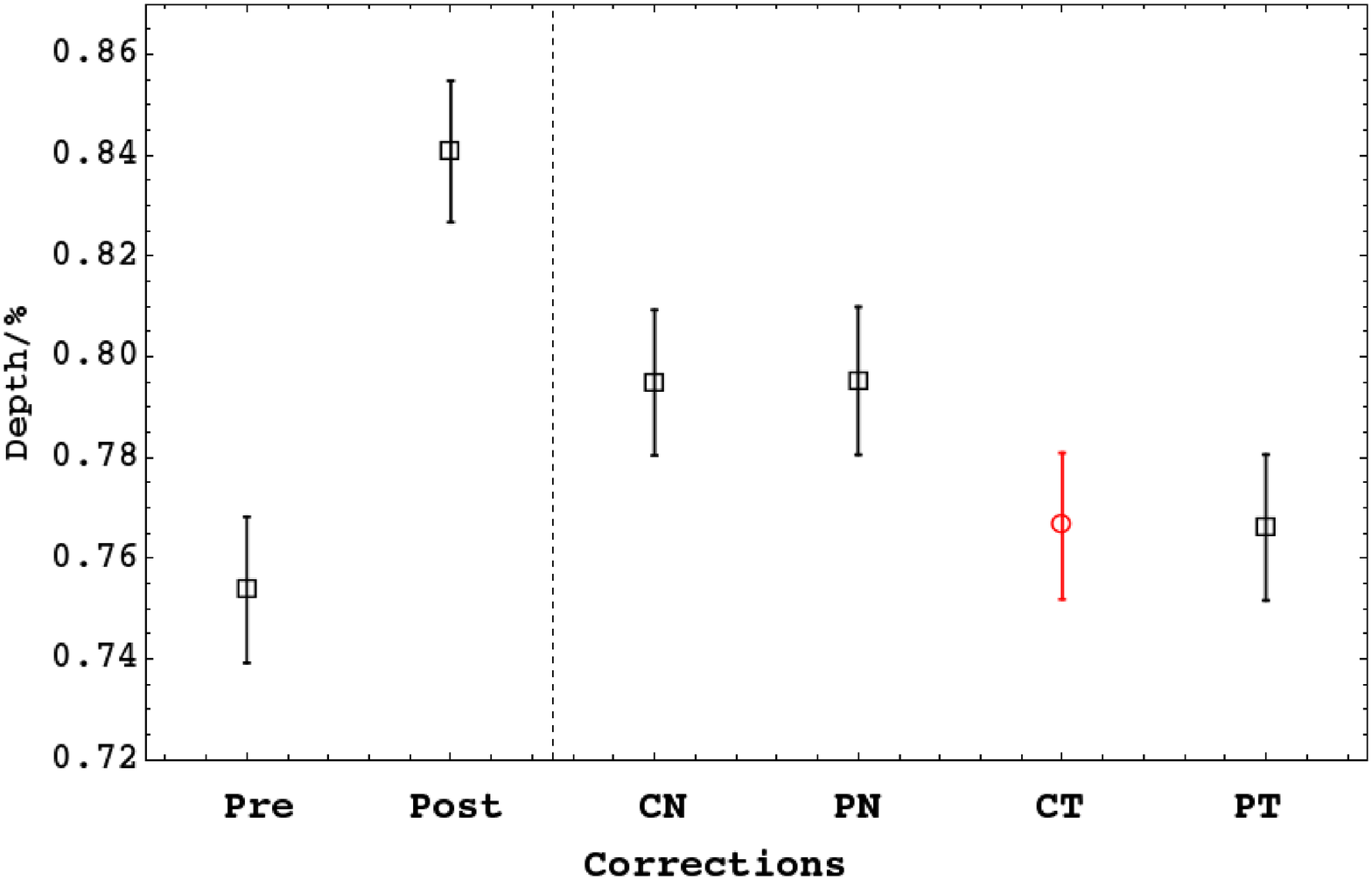} }
\caption{Measured transit depths at 3.6 $\mu$m epoch 1 and 2 (left and right 
upper panels) and 4.5 $\mu$m  (left and right lower panels) respectively, using 
the different families of corrections mentioned in the text. For each family, 
we only show the polynomial order which yielded the lowest Bayesian Information 
Criterion (BIC). The family which yielded the overall lowest BIC is plotted 
using a circle. `Pre' and `Post' are corrections using this same corrective 
function, but using only the pre-transit/post-transit data respectively.}
\label{fig:corr}
\end{center}
\end{figure*}

In this work, we therefore make several different corrections to ensure our 
results are robust. From each family of possible pixel phase corrections, we 
select the order $n$ which produces the lowest BIC, exploring up to quartic 
order. In the final tables, we will only quote the parameters from the absolute 
lowest BIC correction. However, in the different panels of 
Figure~\ref{fig:corr}, we show the transit depths obtained using all four best 
corrections, for comparison. This allows the reader to assess the stability of 
the corrections. Additionally, in these plots, we reproduce the best BIC 
correction but only using i) the pre-transit baseline (`Pre') and/or ii) the 
post-transit baseline (`Post').

As is evident from figure \ref{fig:corr}, the 3.6~$\mu$m results seem very 
stable. 4.5~$\mu$m exhibits some interesting differences. For the first epoch, 
the post-transit only correction leads to a dramatically lower transit depth 
compared to otherwise stable corrections. The reason for this low depth is 
evident when one inspects the residuals of the fit. With barely one cycle of 
pixel phase, the fitted correction is very poor for the pre-transit data. For 
3.6~$\mu$m, making use of just one cycle was probably not such a significant 
hurdle because the signal-to-noise is much higher at 3.6~$\mu$m.

\begin{figure}
\begin{center}
\includegraphics[width=10 cm]{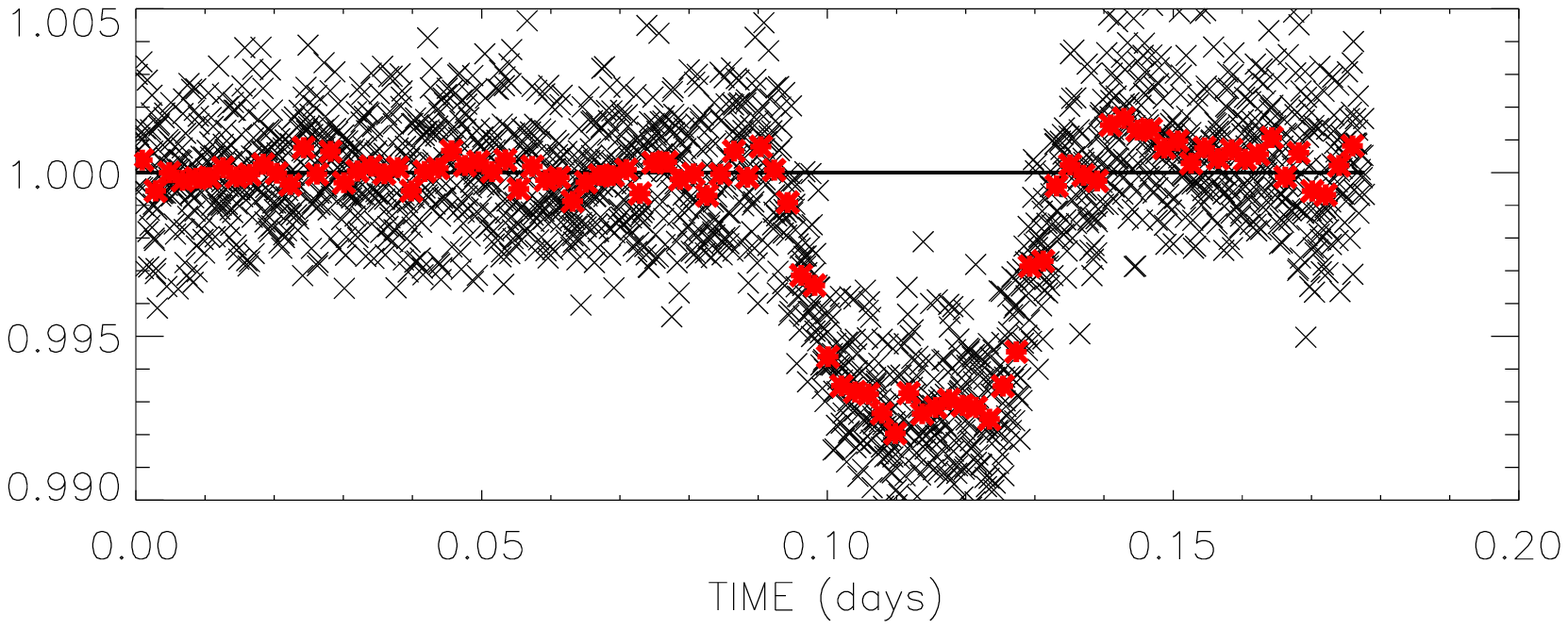}
\caption{We present here the second epoch of 4.5 $\mu$m observations where pixel
phase has been corrected only using pre-transit data. Notice the bump in the 
egress, similar to Stevenson et al. 2010 at 3.6 $\mu$m. We will discard this 
data set for the remainder of the analysis.} \label{fig:ch8precor}
\end{center}
\end{figure}

For the second epoch of 4.5~$\mu$m, the differences are much more severe. The 
pre, post and combined corrections all disagree with one another. 
Correction using  the pre-only transit applied to the full light curve reveals 
flat residuals except for a peculiar bump just after the egress 
(Fig.~\ref{fig:ch8precor}). This bump could be potentially related to the star 
itself and a very similar feature was seen by S2010 for the same target for the 
secondary eclipse at 3.6~$\mu$m. Whilst the hypothesis of stellar induced noise 
is interesting, we have no way of confirming/rejecting such a hypothesis, but 
we note that a tiny change in the pixel phase correction could lead to bumps of 
similar time scale amplitude. We therefore take the pragmatic approach of excluding
 this observation in our final spectral modelling. Since the first epoch 
of 4.5~$\mu$m displays a stable correction, only this value  is used in the 
final modelling.

\subsection{Transit-Systematic phasing}

The systematic trend (in this case the pixel-phase effect) and the transit 
duration exhibit very similar durations and amplitudes. Further, the morphology
of the V-shaped periodic systematic error is quite similar to that of the 
U-shaped transit. We are therefore in a quite undesirable scenario in terms of
systematic correction. The phase difference between the periodic systematic 
error and the transit signal will play a crucial role in the consequences for 
the retrieved transit depth.

Let us define the ``mutual phase'', $\Delta\phi$, between these two signals. We 
define the flux variations caused by the pixel-phase effect alone as 
$F_{\mathrm{systematic}}$. This may be plotted over the uncorrected data. We 
inspect the fitted function around the transit and extract the time stamp of the
minimum in $F_{\mathrm{systematic}}$ both before and after the transit. The 
position of the mid-transit time, $t_C$, between these two limits, is then used
to calculate the mutual phase, $\Delta\phi$. $\Delta\phi = 0^{\circ}$ therefore
indicates that the transit dips down at the exact moment the pixel-phase effect
dips down. We would therefore see an increased apparent transit depth. 
$\Delta\phi = 180^{\circ}$ means the two signals destructively interfere to 
attenuate the apparent transit depth.

\textbf{3.6~$\mu$m}

For the first and second epoch of 3.6~$\mu$m, we obtain 
$\Delta\phi = (0.5 \pm 5.3)^{\circ}$ and $\Delta\phi = (255.9 \pm 4.2)^{\circ}$.
Therefore, for the first epoch, the period, amplitude and phase of the 
systematics effects and the astrophysical signal are almost perfectly aligned.
This is the  worst case scenario  for attempting a correction of  systematic 
effects. As a result of this coincidentally very unfortunate mutual phasing, we
consider the transit depth obtained for the first epoch to be unreliable and 
most probably erroneous. It is therefore not used in our spectral modelling.

A subtle point in this issue is that the pixel phase effect creates a periodic
function with amplitude in the flux direction. There is a very slight drift in 
flux with respect to time, but overall it is in the flux-direction. The transit 
depth is also in the flux direction and therefore will be most severely affected 
by this systematic effect. In contrast, the transit width and ingress duration, 
which affect parameters such $a/R_*$ and impact parameter, $b$, are in the 
time-direction. Although not completely orthogonal, these parameters will be 
much less severely affected by the pixel-phase effect than the transit depth. 
In conclusion,  comparing the derived impact parameter with the known impact 
parameter would be a less reliable method to attempt  to validate the accuracy 
of a pixel-phase correction.

\textbf{4.5~$\mu$m}

For 4.5~$\mu$m, the phase angles for the first and second epoch are 
$\Delta\phi = (159.4 \pm 4.5)^{\circ}$ and 
$\Delta\phi = (303.0 \pm 4.4)^{\circ}$.  As we have already discussed, the 
second epoch of 4.5~$\mu$m was not  used anyway because the transit depth was 
shown to be unstable with respect to the pixel phase correction used. Therefore,
we only need consider the first epoch, which does not exhibit an extreme mutual 
phasing and thus should be reliable.

We note that the optimum strategy would be to have numerous transit observations
at different values $\Delta\phi$ which fully span the region 
$0^{\circ}\leq\Delta\phi<360^{\circ}$. By analysing all of the transit depths 
at each angle, the effect of mutual phasing could potentially be removed giving
a more reliable estimate of the transit depth. However, we do not have a large 
number of transit measurements and so we are forced to proceed in the most 
reasonable way possible with the limited data presently available.

\section{Ramp correction at 8 $\mu$m}

\begin{figure}
\begin{center}
\includegraphics[angle=0,width=10 cm]{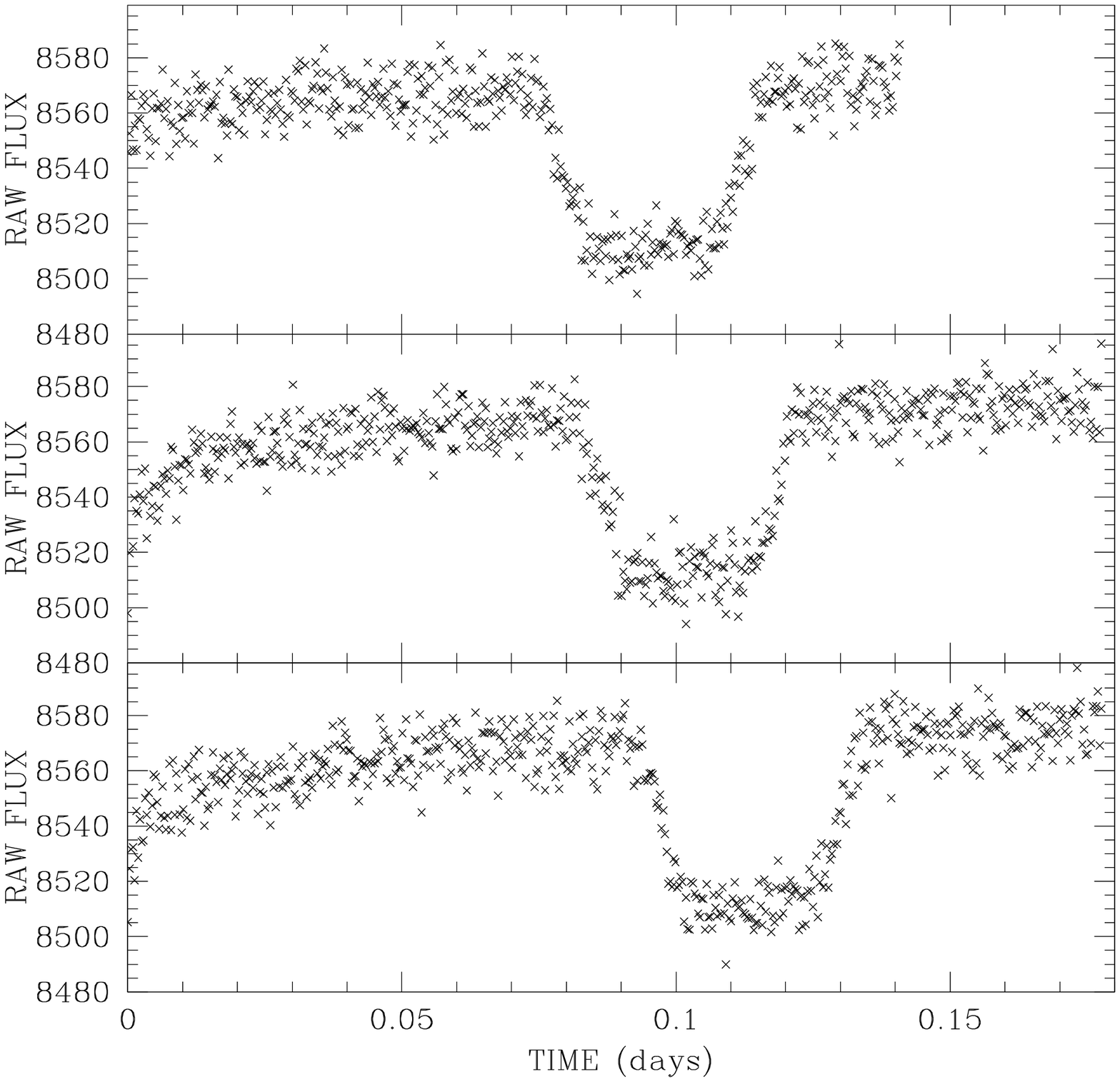} 
\caption{ Raw photometric data for the three epochs at 8 $\mu$m obtained with IRAC. 
.} \label{fig:lc8m}
\end{center}
\end{figure}

There is a variation of the response of the pixels to a long period of 
illumination and latent build-up effect 
impinges on the 8.0 $\mu$m observations, called ``the ramp'', whilst their pixel
phase effects are negligible. We show the three raw light curves in 
Figure~\ref{fig:lc8m}. { We will perform two fits to correct the data. 
First, following Agol et al. (2009) and Beaulieu et al. (2010) we fit the 
$f(a,b,c,d,t_0,t)=a+b (t-t_0)+c  \log(t-t_0)$ function to the out 
of transit data.  The second approach recently presented by Agol et al. (2010) 
adopts a function of the form 
$f(a_0,a_1,a_2,\tau_1,\tau_2,t_0,t) = a_0-a_1 e^{-(t-t_0)/tau_1 } -a_2 e^{-(t-t_0)/tau_2 } $ }. 
Data corrected for systematics are shown in Figure 3. { The two 
correction are not distinguishable by eye and the measured transits depth agrees within
the error bars.  Moreover our results } are compatible with Deming et al. (2007) 
who reported $0.704 \pm 0.009\%$ at 8~$\mu$m. {In the final quoted 
values, we opt for the simpler model of Agol et al. (2009) which has just three 
fitted parameters. Final values for the ramp parameters are given in
Table~\ref{table:rampparams}.}

For each of three epochs, the corrected transit lightcurve is fitted 
independently using the same fitting parameter set used in 
Kipping \& Bakos (2010) and Kipping (2010), method A, namely 
\{$t_c$,$p^2$,$\Upsilon/R_*$,$b^2$,$OOT$\}. The derived parameters are listed in
Table 1. Let us define our null hypothesis to be that the transit depths 
obtained for 8.0~$\mu$m are consistent with being sampled from the same Gaussian
distribution. The scatter in the depths suggests that this hypothesis is 
rejected at the 1.4-$\sigma$ level, which is not statistically significant. We
therefore conclude that there is no evidence to support a hypothesis of transit
depth variation or any evidence that our systematic corrective procedure has 
introduced biases into any of the depths. Given this conclusion, a global fit of
all three 8.0~$\mu$m measurements is justified. Fitting for the period as an 
extra free parameter, we collectively fit all three lightcurves, giving the 
result displayed in table 1.

\begin{table*}
{\tiny
\begin{tabular}{l l l} 
\hline 
Parameter & Best fit & Standard Error  \\
\hline
\multicolumn{3}{c}{ {\it {\bf Channel 4}} }\\
\hline
$a$ & 8539.1 & 5.5 \\
$b$ & 0.00012 & 0.00022 \\
$c$ & 3.21 & 0.78 \\
\hline
\multicolumn{3}{c}{ {\it {\bf Channel 6}} }\\
\hline 
$a$ & 8487.3 & 5.0 \\
$b$ & -0.00012 & 0.00015 \\
$c$ & 9.33 & 0.69 \\
\hline  
 \multicolumn{3}{c}{ {\it {\bf Channel 7}} }\\ 
\hline  
$a$ & 8508.7 & 4.6 \\
$b$ & 0.00002 & 0.00014 \\
$c$ & 6.73 & 0.65 \\
\hline 
\end{tabular} }
\caption{\emph{Best-fitted ramp correction parameters for each 8\,$\mu$m time 
series. $t_0$ was selected to be $-30$\,s. Best-fit parameters computed using a
Levenberg-Marquardt algorithm. All errors quoted to 2 significant figures
and corresponding best-fit values to the equivalent number of decimal places.}} 
\label{table:rampparams} 
\end{table*}

\section{Comments on Spitzer secondary transits observations by Stevenson et al.}

We performed our own photometry and analysis of the 8-$\mu$m secondary
transit data used in S2010 and obtained identical results both for the
transit depth and its uncertainty. We also
reprocessed the data at 5.8 $\mu$m, correcting for systematics, and
measured a secondary-transit depth of $0.036 \pm 0.017 \%$, compatible
with the S2010 results.

\begin{figure}
\begin{center}
\includegraphics[width=10 cm]{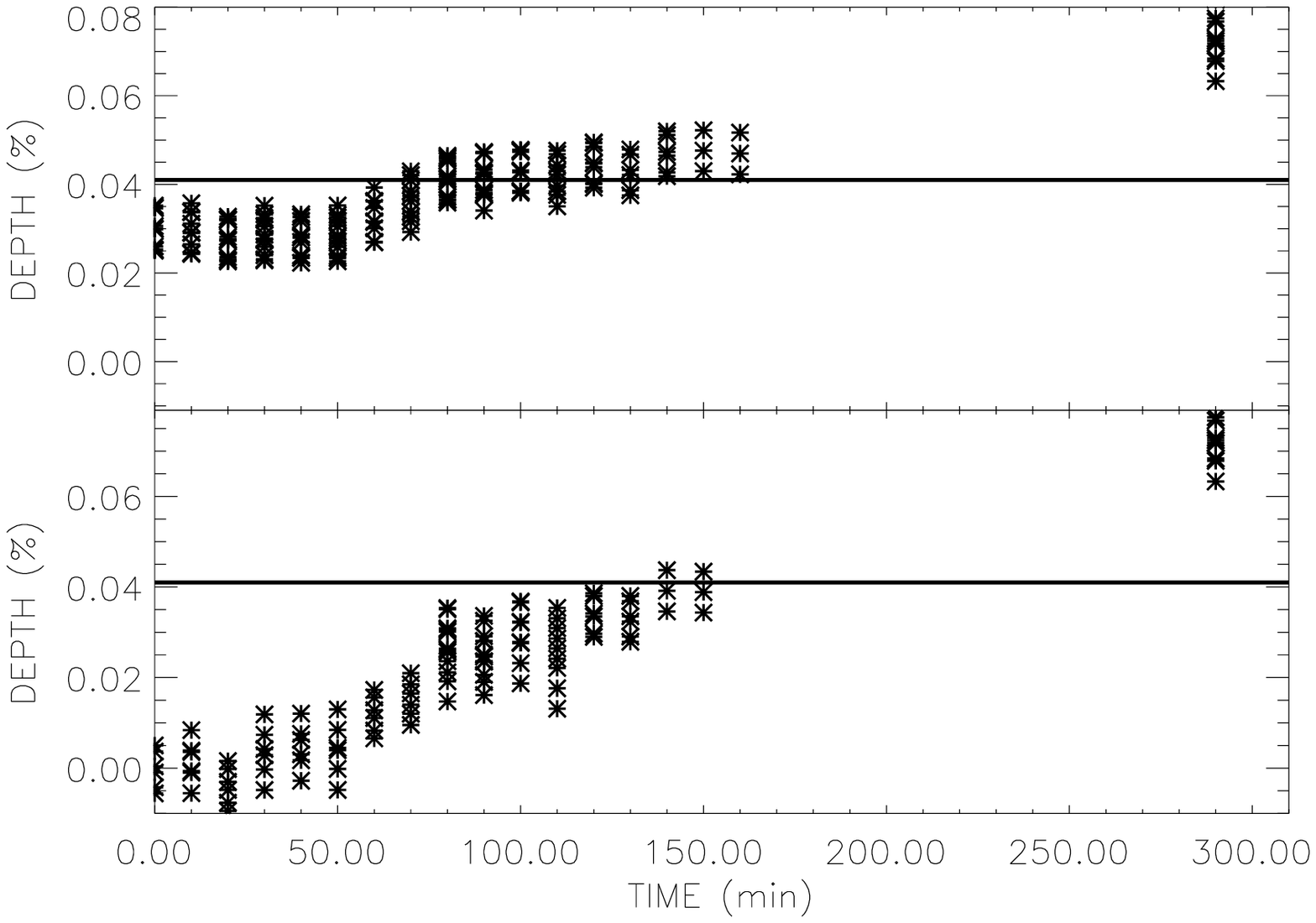}
\caption{We present here measurements of the secondary transit depth at 3.6 $\mu$m as a function of the
start time of the intransit observations for different length of time to derive the corrective terms for systematics.
See the text for detail about the procedure.} \label{fig:C1}
\end{center}
\end{figure}

We have re-examined the 3.6-$\mu$m and 
4.5-$\mu$m observations using similar procedures to those described here
for primary transits. For the 3.6-$\mu$m data the best correction  is obtained
by using quadratic terms for pixel phase and no time dependence, giving
a secondary transit depth of $0.041 \pm 0.006 \%$. 
This value is identical to S2010, but the uncertainty is twice as large. However, we also
note that the post-transit spike, reported in S2010 and noted as 
being possibly due to stellar activity, may bias 
the pixel-phase correction, as was also found here for 4.5-$\mu$m data in
primary transit. As a check, we decided
to exclude the 3.6-$\mu$m post-transit data and spike, the first 70 minutes of observations
and recomputed the pixel-phase correction and eclipse depth: we obtain a
secondary transit depth of $0.02 \pm 0.006 \%$
These two results are incompatible and do
not incorporate an error from the systematics.
 The two results do not  incorporate any systematic uncertainty from the pixel-phase correction,
but are formally inconsistent. 
{ We then decided to explore the sensitivity to the chosen section of out of transit data
used to estimate for the pixel phase effect to correct the light curve. 
First, we considered only pre-transit data to estimate the correction. We took sections
of data of 80 minutes, 100 minutes, 120 minutes, 140 minutes, 160 minutes with different start
time. We then compute the correction and fit the transit depth.  We perform similar measurement 
too for the post transit observations alone. Results are reported in the lower pannel of Figure \ref{fig:C1}. 
We incorporated post transit between 290 and 310 minutes after the first exposure and performed 
the same corrections and fits. It is clear that depending on the section of the data that is
used to derive the correction different results are obtained, and that the transit depth measurement
is completely dominated by systematic errors that are not under control.
Results are shown in the upper pannel of Figure \ref{fig:C1}. We consider 
that no reliable measurement could be obtained for this epoch. We also would like to comment
about the post spike seen from S2010 and also observed in our different re-reduction. Rather than
a photometric variation timed at the end of the secondary transit, we suggest that it could also be
remaining systematics of the instrument similar to what we observed in Figure A3.}

We also estimate the mutual phase between 
transit and pixel-phase to be $\Delta\phi = (5 \pm 10)^{\circ}$. 
 This close mutual  phasing and the presence of the post-transit spike would normally lead us to 
exclude the 3.6-$\mu$m dataset from further analysis; indeed, this highlights
the acute need for further observations of the secondary transit of GJ436b at
3.6 $\mu$m at different mutual phases in order to check this transit depth, 
especially given the high leverage exerted by this point in the S2010 
analysis of the degree of methane absorption in secondary transit. We will adopt a { conservative } $0.03 \pm 0.02 \%$.

We also performed the secondary-transit analysis of 4.5 $\mu$m data and 
find a measured transit depth of $0.01 \pm 0.01 \%$, while S2010 found a 3-sigma upper limit at $0.01 \%$. 
We conclude that the two main discrepancies between S2010 and our own 
analysis are at 3.6 and 4.5~$\mu$m, {\em i.e.}, those points which carry the
greatest weight in their analysis and conclusions. { Measurements at 3.6 $\mu$m and 4.5 $\mu$m  should be 
  redone with warm Spitzer. We encourage further studies oh transiting exoplanets 
to obtain multiple epoch for transit measurements, in particular when  the transit duration is of the 
same time scale as the pixel scale effects such as for COROT 7b, GJ1214b and GJ436b.  }



\end{document}